\documentclass{aa}  

\usepackage{graphicx}
\usepackage{xcolor}
\usepackage{txfonts}
\begin{document}

   \title{PLATOSpec's first results: Three new transiting warm Jupiters from the WINE survey TIC\,147027702, TIC\,245076932 and TIC\,87422071}

    \author{Pavol Gajdo\v{s}
      \inst{1,2}
      \and
      Rafael Brahm\inst{3,4}
      \and 
      Lorena Acu\~{n}a-Aguirre\inst{5}
      \and
      Mat\'{i}as I. Jones\inst{6}
      \and
      Helem Salinas\inst{3}
      \and 
      Jozef Lipt\'{a}k
      \inst{1,7}
      \and
      Andrés Jordán\inst{3,4}
      \and
      Thomas Henning\inst{5}
      \and
      Ji\v{r}\'{i} Srba 
      \inst{1}
      \and 
      Eva \v{Z}\v{d}\'{a}rsk\'{a} 
      \inst{1}
      \and 
      Zuzana Balk\'{o}ov\'{a} 
      \inst{1,7}
      \and 
      Michaela V\'{i}tkov\'{a} 
      \inst{1,8}
      \and
      Jan Jan\'{i}k
      \inst{8}
      \and
      Petr \v{S}koda
      \inst{1}
      \and
      Ji\v{r}\'{i} \v{Z}\'{a}k
      \inst{1}
      \and
      Djamel Mekarnia
      \inst{9}
      \and
      Olga Suarez  
      \inst{9}
      \and
      Lyu Abe
      \inst{9}
      \and
      Matteo Beltrame
      \inst{10,11}
      \and
      Amaury H.M.J. Triaud
      \inst{12}
      \and
      Tristan Guillot
      \inst{9}
      \and
      Karen A.\ Collins
      \inst{13}
      \and
      Khalid Barkaoui
      \inst{14,15,16}
      \and
      Gavin Boyle
      \inst{17,18}
      \and
      Vincent Suc
      \inst{4,3,17}
      \and
      Luca Antonucci
      \inst{19,20}
      \and
      Marcelo Tala Pinto
      \inst{3,4}
      \and
      Elizaveta Vostretcova
      \inst{21}  
      \and
      Jan Eberhardt
      \inst{5}
      \and
      N\'estor Espinoza
      \inst{22}
      \and
      Ismael Mireles
      \inst{23}
      \and
      Pavel Pintr
      \inst{24} 
      \and
      Felipe I. Rojas
      \inst{25}
      \and
      Veronika Schaffenroth
      \inst{21}
      \and
      Leo Vanzi
      \inst{19,20}
      \and 
      Petr Kab\'{a}th
      \inst{1}
      }
      \authorrunning{Gajdo\v{s} et al.}
      \titlerunning{Three new transiting warm Jupiters}

   \institute{Astronomical Institute, Czech Academy of Sciences, Fri\v{c}ova 298, 25165 Ond\v{r}ejov, Czech Republic\\
              \email{gajdos@asu.cas.cz}
         \and
             Institute of Physics, Faculty of Science, Pavol Jozef \v{S}af\'arik University, Park Angelinum 9, 04001 Ko\v{s}ice, Slovakia
             \email{pavol.gajdos@upjs.sk}
        \and
            Facultad de Ingeniería y Ciencias, Universidad Adolfo Ibáñez, Av. Diagonal Las Torres 2640, Peñalolén, Santiago, Chile
        \and
            Millennium Institute for Astrophysics, Av. Vicuna Mackenna 4860, 782-0436 Macul, Santiago, Chile
        \and
            Max-Planck-Institut f\"{u}r Astronomie, K\"{o}nigstuhl 17, D-69117 Heidelberg, Germany
        \and
        European Southern Observatory (ESO), Alonso de C\'ordova 3107, Vitacura, Casilla 19001, Santiago, Chile
        \and
        	Astronomical Institute of Charles University, V Hole\v{s}ovi\v{c}k\'{a}ch 2, CZ-180 00 Prague, Czech Republic
       \and       
            Department of Theoretical Physics and Astrophysics, Faculty of Science, Masaryk University, Kotl\'{a}\v{r}sk\'{a} 2, CZ-611 37 Brno, Czech Republic
        \and 
        Université Côte d'Azur, Observatoire de la Côte d'Azur, CNRS, Laboratoire Lagrange, CS 34229, F-06304 Nice Cedex 4, France
        \and
        Istituto di Scienze Polari del CNR (ISP-CNR), Università Ca’ Foscari, Via Torino n. 155, 30172 Venezia Mestre (VE), Italy
        \and
        Programma Nazionale di Ricerche in Antartide (PNRA),Institut polaire français Paul-Émile Victor (IPEV)
        \and
        School of Physics \& Astronomy, University of Birmingham, Edgbaston, Birmingham B15 2TT, UK
        \and
        Center for Astrophysics \textbar \ Harvard \& Smithsonian, 60 Garden Street, Cambridge, MA 02138, USA
        \and 
        Instituto de Astrof\'isica de Canarias (IAC), Calle V\'ia L\'actea s/n, 38200, La Laguna, Tenerife, Spain
        \and
        Astrobiology Research Unit, Universit\'e de Li\`ege, 19C All\'ee du 6 Ao\^ut, 4000 Li\`ege, Belgium
        \and
        Department of Earth, Atmospheric and Planetary Science, Massachusetts Institute of Technology, 77 Massachusetts Avenue, Cambridge, MA 02139, USA
        \and   
        El Sauce Observatory --- Obstech, Coquimbo, Chile
        \and
        Cavendish Laboratory, J. J. Thomson Avenue, Cambridge, CB3 0HE, UK
        \and
        Center of Astro Engineering, Pontificia Universidad Católica de Chile, Av. Vicuña Mackenna 4860, 782-043 Santiago, Chile
        \and
        Department of Electrical Engineering, Pontificia Universidad Católica de Chile, Av. Vicuña Mackenna 4860, 782-043 Santiago, Chile
        \and
        Thüringer Landessternwarte, D-07778 Tautenburg, Germany
        \and
        Space Telescope Science Institute, 3700 San Martin Drive, Baltimore, MD 21218, USA
        \and
        Department of Physics and Astronomy, The University of New Mexico, Albuquerque, NM 87106, USA
        \and
        Institute of Plasma Physics of the Czech Academy of Sciences, Research Centre for Special Optics and Optoelectronic Systems TOPTEC, U Slovanky 2525/1a 182 00, Praha 8
        \and
        Instituto de Astrof\'isica, Pontificia Universidad Cat\'olica de Chile, Av. Vicu\~na Mackenna 4860, 7820436 Macul, Santiago, Chile
                }

   \date{Received ; accepted}

 
\abstract{We report the discovery and characterisation of three transiting warm Jupiters: TIC\,147027702b, TIC\,245076932b and TIC\,87422071b. These systems were initially identified as transiting candidates using light curves generated from the full-frame images of the TESS mission. We confirmed the planetary nature of these objects with ground-based spectroscopic follow-up observations using FEROS and the new PLATOSpec spectrograph attached to the ESO 1.52~m telescope at the La Silla Observatory, and with ground-based photometric observations of the Observatoire Moana, Las Cumbres Observatory Global Telescope and ASTEP. From a global fit to the photometry and radial velocities, we determine that the planet TIC\,147027702b has a low-eccentric orbit ($e = 0.13 \pm 0.05$) with a period of 44.4 days and has a mass of $1.09^{+0.07}_{-0.13}$~M$_J$ and a radius of $0.98 \pm 0.06$~R$_J$. TIC\,245076932b has a moderately low mass of $0.51 \pm 0.05$~M$_J$, a radius of $0.97 \pm 0.05$~R$_J$, and an eccentric orbit ($e = 0.43 \pm 0.02$) with a period of 21.6 days. TIC\,87422071b has a mass of $1.29 \pm 0.10$~M$_J$, a radius of $0.97 \pm 0.08$~R$_J$, and has a slightly eccentric orbit ($e = 0.12 \pm 0.07$) with a period of 11.3 days. These well-characterised warm Jupiters expand the currently limited sample of similar gas giants and provide valuable benchmarks for testing models of giant-planet formation, migration, and tidal evolution.
   }

   \keywords{Techniques: photometric -- Techniques: radial velocities -- planetary systems -- Planets and satellites: gaseous planets -- Planets and satellites: detection\vspace{-1.1cm}}

   \maketitle

\section{Introduction}

The number of discovered exoplanets has increased over the years and currently exceeds 6,000. The majority of well-characterised giant planets are hot Jupiters - highly irradiated gas planets at extreme proximities from their host stars \citep[e.g.][]{Santerne2016}. These planets are not expected to form at their current locations, given that the formation of planetary cores is highly inefficient inside the snowline  \citep[e.g.][]{Rafikov2005, Schlichting2014}. A possible mechanism to form the hot Jupiter population is migration from beyond the snowline, either by interactions with the gaseous disc when it is still present or through high eccentricity tidal migration mechanisms, generated by interactions with other objects in the system after the dispersal of the disc \citep[e.g.][]{Kley2012,Walsh2011}.

Warm giant planets could be perceived as a transition state in the evolution of hot Jupiters. Warm Jupiters are generally defined as planets with masses greater than 0.3 M$_{\rm J}$ and orbital periods between 10 and 200 days \citep{Dong2021}. Their equilibrium temperatures are below $\sim 1000$~K. The observed properties, planetary atmospheres, and physical processes driving the structure of warm Jupiters are less affected by proximity to the host star, which makes them a unique opportunity to test planetary system formation and evolution theories. However, the number of discovered and well-characterised transiting warm giant planets is small -- up to today, about 100\footnote{Data from NASA Exoplanet Archive}.

Here, we present the discovery, confirmation, and characterisation of three new transiting planets belonging to the class of warm Jupiters. They were identified as planetary candidates from the TESS light curves \citep{tess} and confirmed by follow-up radial velocity measurement using the PLATOSpec spectrograph \citep{Kabath2025}. In addition, we performed ground-based photometric observations to confirm the true source of the transit signal.

This paper is organised as follows. We describe the photometric and spectroscopic observations used in Sect.~\ref{obs}. Sec.~\ref{star} presents the stellar parameters of the parent star. In Sect.~\ref{model}, we provide the parameters of the planetary system obtained by our analysis. Finally, we discuss our results and draw conclusions in Sect.~\ref{con}.

\section{Observations}
\label{obs}

\subsection{TESS photometry}

\begin{figure*}[ht]
\begin{center}
\includegraphics[width=1.0\linewidth]{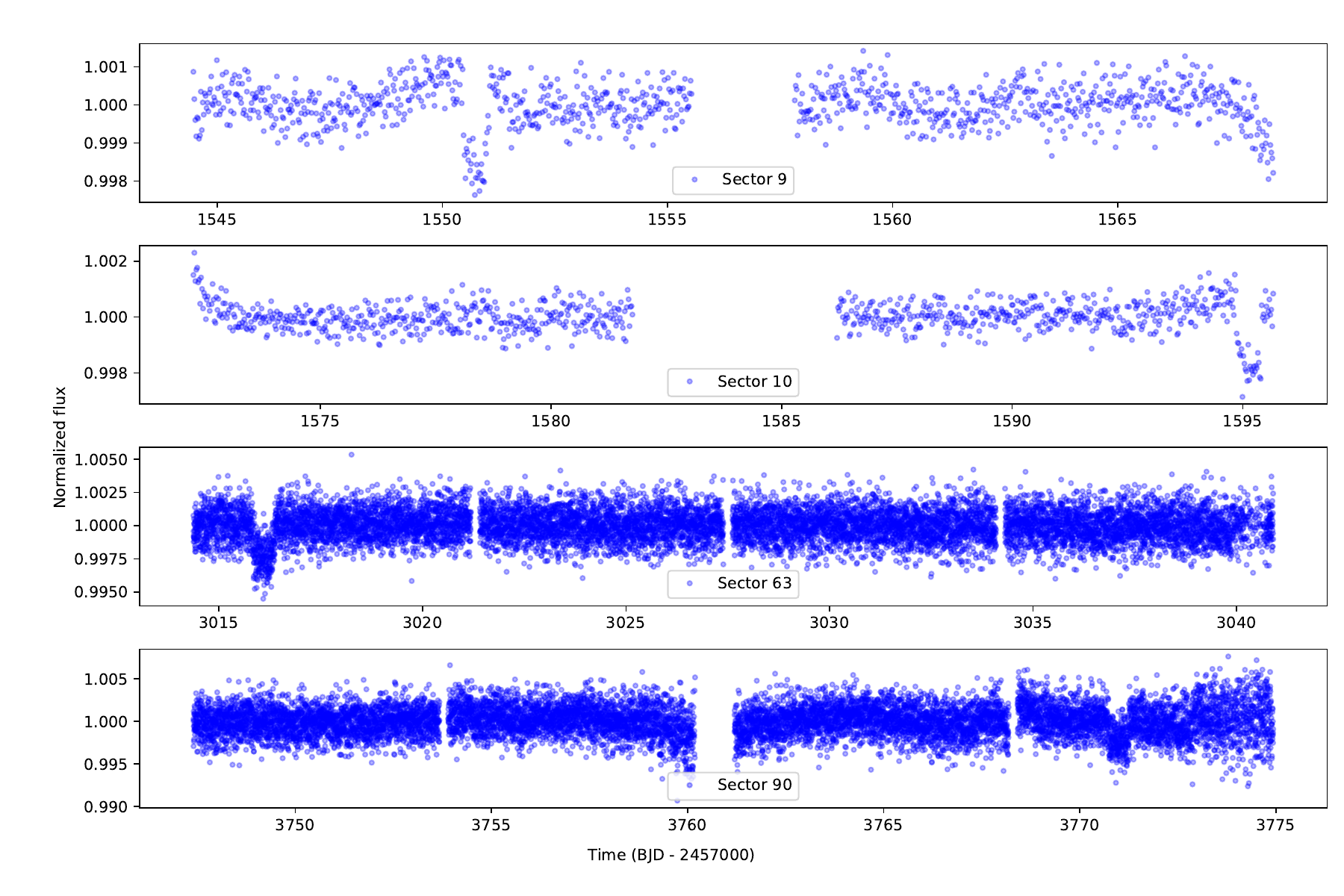}
\end{center}
\caption{\textit{TESS} photometry of four sectors used in our analysis of TIC\,147027702.}
\label{fig:tic147_tess-all}   
\end{figure*}

TIC\,147027702b was identified as a transiting planetary candidate from the full-frame image (FFI) light curves of the TESS mission using a transformer-based algorithm \citep{Salinas2025} in the context of the Warm gIaNts with tEss collaboration \citep[WINE;][]{brahm:2019,jordan:2020,schlecker:2020,trifonov:2021,hobson:2023,eberhardt:2023,jones:2024,tala:2025}. The candidate was marked as a single transiter based on the observation from Sector 9 (March 2019). A posterior analysis revealed three additional transits in Sectors 10 (April 2019), 63 (March 2023), and 90 (March 2025). This target was also observed by TESS in Sector 36 (March 2021), but no transit occurred during that period. For the first two sectors (S9 and S10), the FFIs were collected with a cadence of 30 min. In Sectors 63 and 90, the cadence was changed to 200 seconds. The time series of these sectors are shown in Fig.~\ref{fig:tic147_tess-all}. From these four TESS light curves, it was possible to constrain the orbital period of the transiting candidate to P$\approx$44.4 d. The transit depth of $\delta \approx 2200$\ ppm, and therefore, is consistent with a possible warm giant planet.

\begin{figure*}[ht!]
\begin{center}
\includegraphics[width=0.3\linewidth]{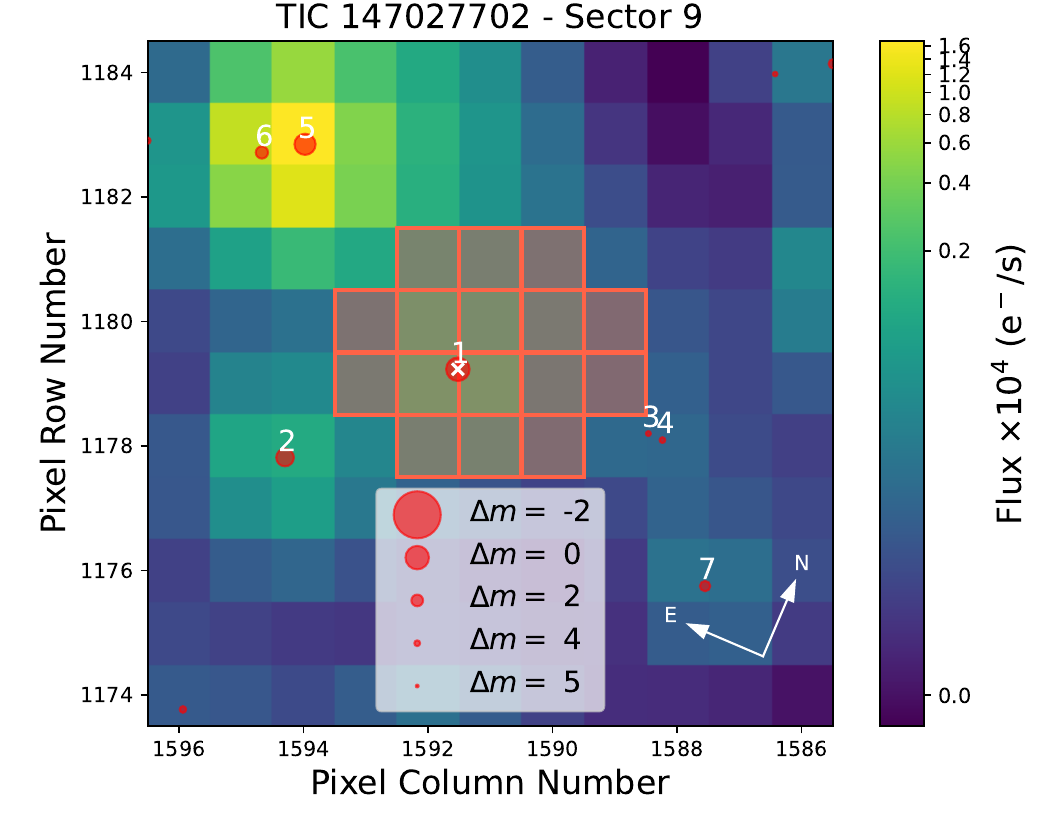}
\includegraphics[width=0.3\linewidth]{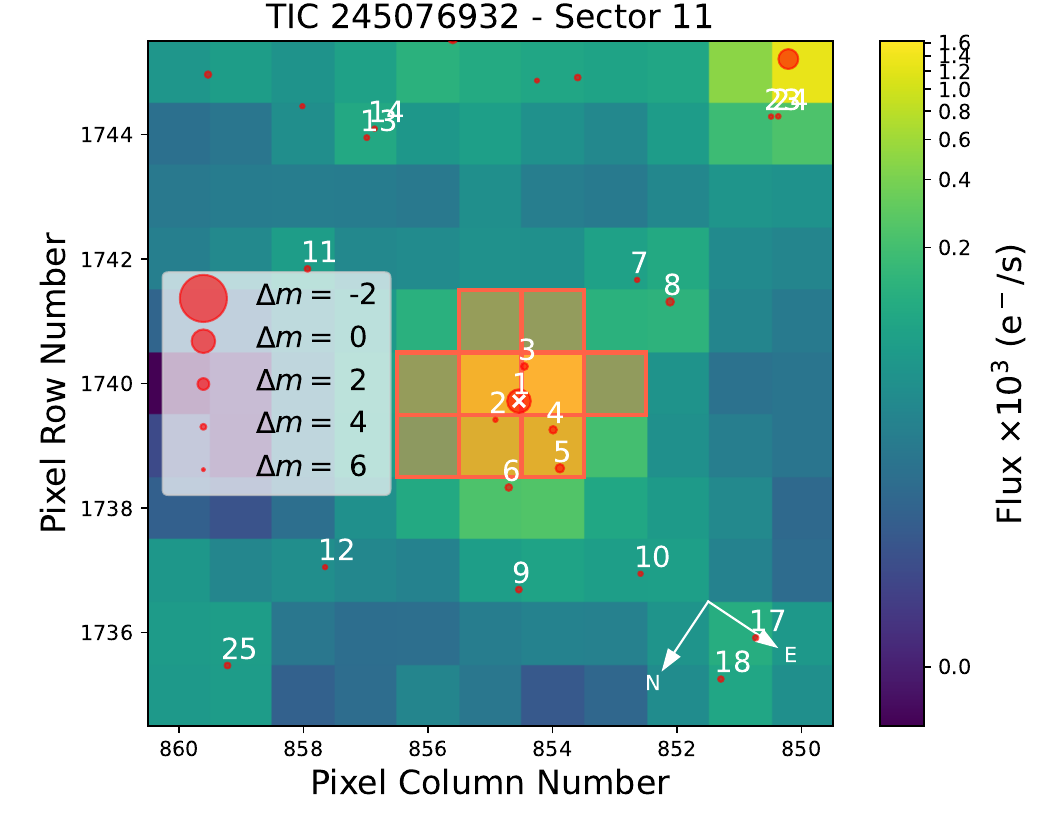}
\includegraphics[width=0.3\linewidth]{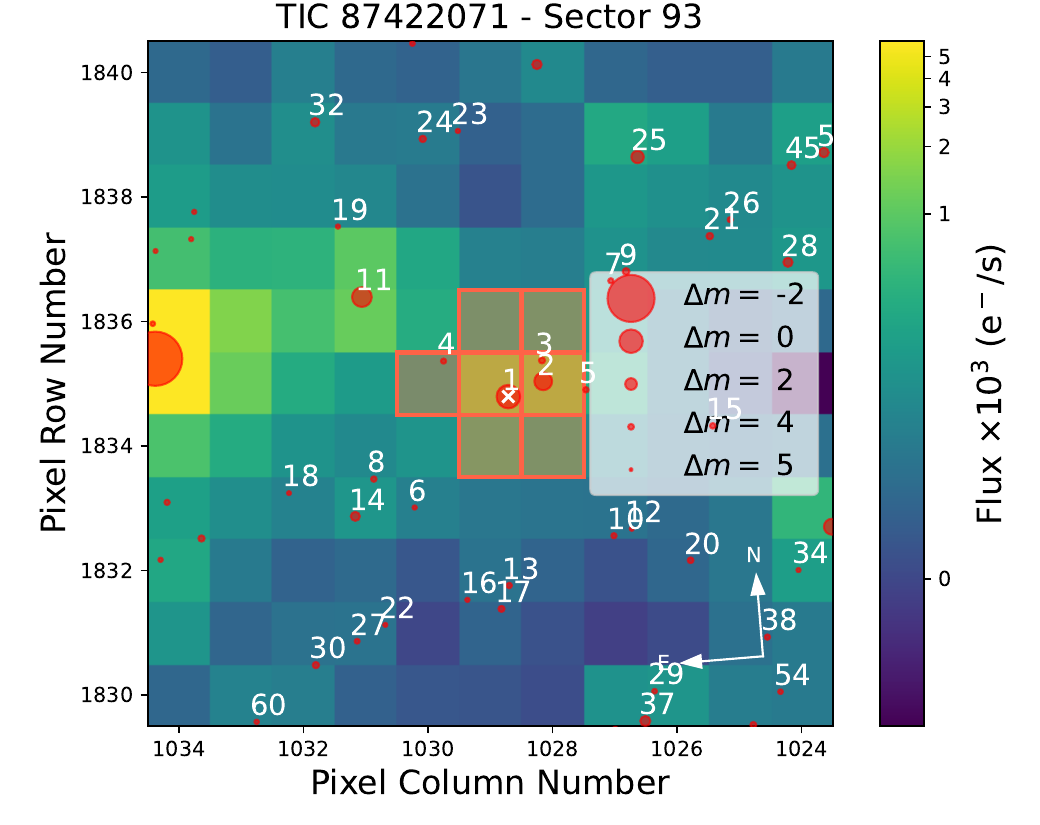}
\end{center}
\caption{\textit{TESS} target pixel file (TPF) of TIC\,147027702 (\textit{1$^{\rm st}$ panel}), TIC\,245076932 (\textit{2$^{\rm nd}$ panel}) and TIC\,87422071 (\textit{3$^{\rm rd}$ panel}).}
\label{fig:tpf}
\end{figure*}

\begin{figure*}[ht]
\begin{center}
\includegraphics[width=1.0\linewidth]{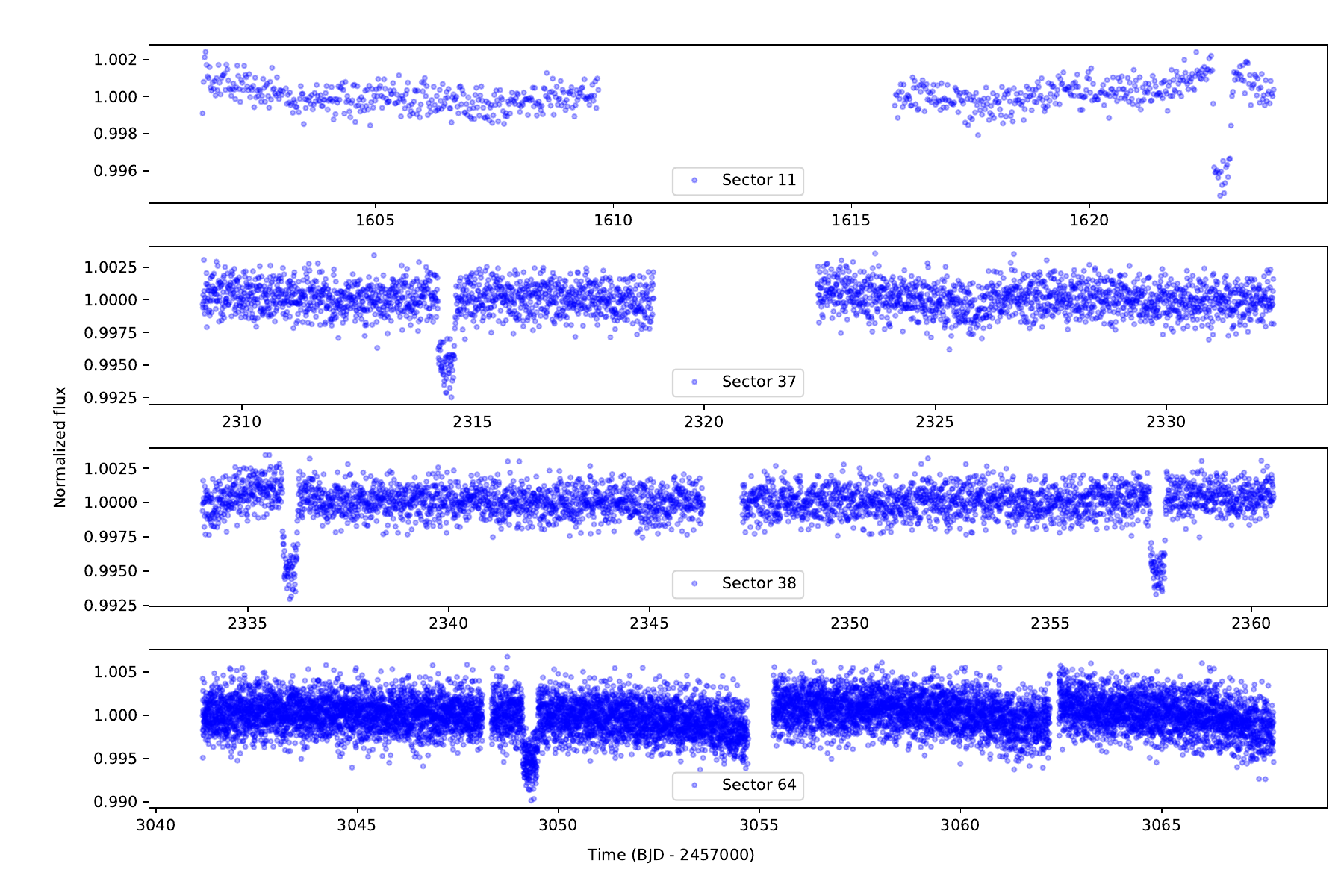}
\end{center}
\caption{\textit{TESS} photometry of four sectors used in our analysis of TIC\,245076932.}
\label{fig:tic245_tess-all}   
\end{figure*}

\begin{figure*}[ht]
\begin{center}
\includegraphics[width=1.0\linewidth]{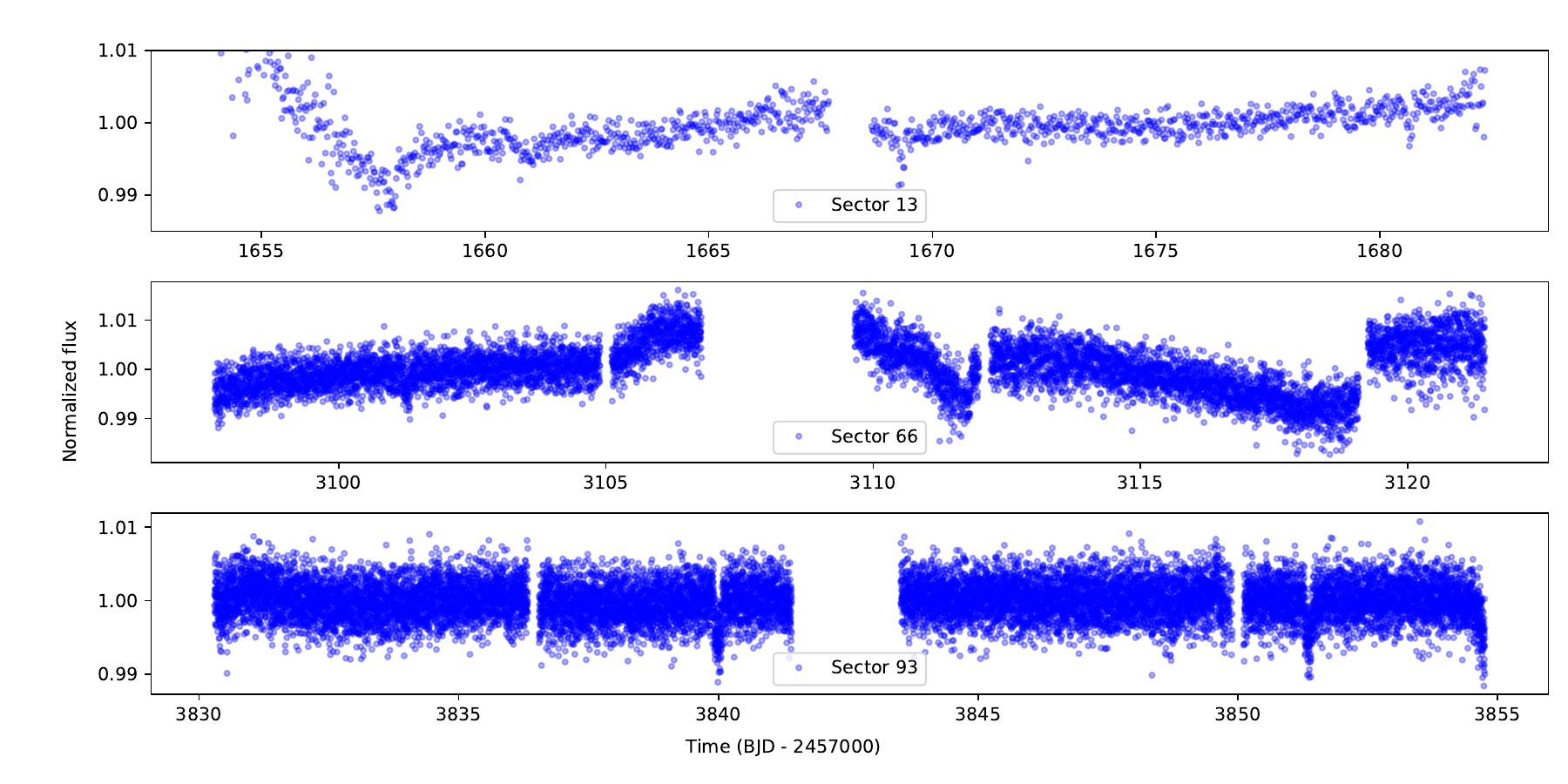}
\end{center}
\caption{\textit{TESS} photometry of three sectors used in our analysis of TIC\,87422071.}
\label{fig:tic874_tess-all}   
\end{figure*}

Using a similar approach, a new transiting candidate was identified on the star TIC\,245076932. However, multiple transits were detected in the light curves of the TESS primary mission, and the orbital period of the planetary candidate was estimated to be 21.6 days by \citet{Salinas2025}. TESS observed this target during four sectors: Sector 11 (May 2019), Sector 37 (April 2021), Sector 38 (May 2021), and Sector 64 (April 2023). The cadence of FFIs in Sector 11 was 30 min. It was changed to 10 min in Sectors 37 and 38 and to 200~s in Sector 64. From their time series (Fig.~\ref{fig:tic245_tess-all}), we identified 5 full transits with a depth of $\delta \approx 5990$\ ppm. An additional partial transit was observed in Sector 11, where only the end of the engres is captured.

TIC\,87422071 was identified as a planetary candidate with an orbital period of 11.36 days from the TESS quick-look pipeline \citep[QLP,][]{qlp}.
Seven transits in three TESS sectors (Fig.~\ref{fig:tic874_tess-all}) were identified with a depth of $\delta \approx 4470$\ ppm. Sector 13 (July 2019) has an FFIs cadence of 30 min. An additional two sectors, Sector 66 (June 2023) and Sector 93 (June 2025), were taken with a 200-s cadence.

Some additional out-of-transit variability can be observed in the LCs for each target, which varies between individual sectors. It is mainly the result of systematic trends in TESS data. There are significant differences in the amount of detrending comparing the used data product (QLP or SPOC). The SPOC pipeline eliminates most of the additional trends. Stellar variability (e.g. pulsations or stellar spots) has probably less impact in our case.

We analyse the target-pixel files (TPFs) for possible contamination from other sources. We have used \texttt{tpfplotter} \citep{Aller2020} for this purpose. Figure~\ref{fig:tpf} shows the TPF for all three targets. There are no neighbouring stars inside the aperture mask of TIC\,147027702. Only a few fainter stars (with a brightness difference of more than 2 mag) were identified in the background with an angular separation from the target of at least 1.5\arcmin. Inside the aperture mask of TIC\,245076932, multiple faint stars can be identified. However, the brightness difference is 3 to 5 mag. TIC\,87422071 has a close bright neighbour. The star TIC\,87422067 (Gaia DR3 6721152024675106816) is separated by only 12" with a $V$ mag of 12.82 (compared to 11.98 of our target). Ground-based photometry was therefore necessary to prove the real source of the transiting signal. The additional three stars fainter than 15.5 mag are presented in its aperture mask. 

\subsection{Ground-based photometry}

We acquired ground-based photometric observations to confirm the true source of the transit signal. For this purpose, we used multiple telescopes in various observatories.

Observatoire Moana (OM) is a global network of small-aperture robotic optical telescopes. The station installed at the EL Sauce observatory in Chile consists of a 0.6 m Dall-Kirkham robotic telescope coupled to an Andor iKon-L 936 deep depletion $2k\times2k$ CCD with a scale of $0.67 \arcsec$ per pixel.

ASTEP (Antarctica Search for Transiting ExoPlanets) is a 40 cm telescope installed at the Concordia station on the Antarctic plateau \citep{Dransfield2022,astep,Schmider2022}. The FLI ProLine PL16801 $4096\times4096$ CCD camera has a field of view of $1\degr \times 1\degr$ and a scale of $0.92\arcsec$ per px.

The Las Cumbres Observatory Global Telescope (LCOGT) \citep{Brown:2013} 1.0\.m network node at South Africa Astronomical Observatory near Sutherland (SAAO) is equipped with a $4096\times4096$ SINISTRO camera having an image scale of $0.389\arcsec$ per pixel, resulting in a $26\arcmin\times26\arcmin$ field of view.

\begin{figure*}[h]
\begin{center}
\includegraphics[width=0.3\linewidth]{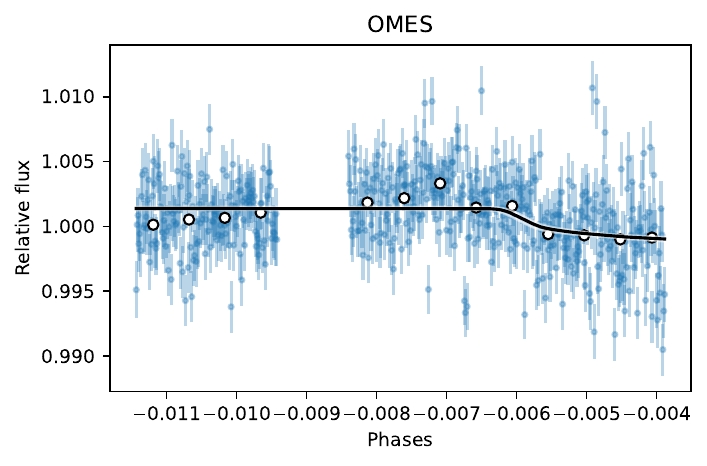}
\includegraphics[width=0.3\linewidth]{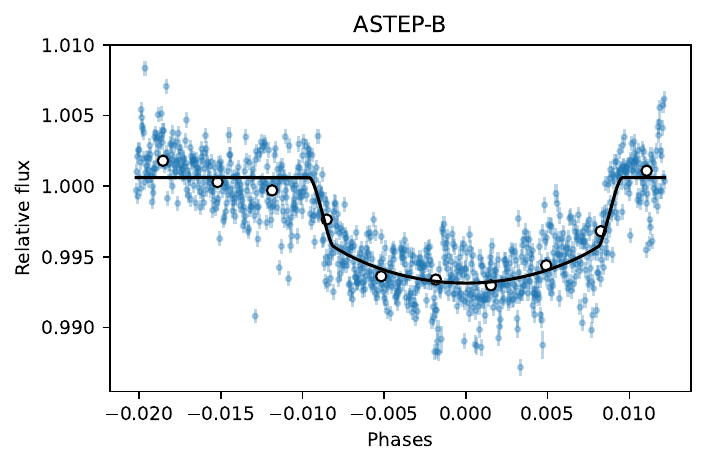}
\includegraphics[width=0.3\linewidth]{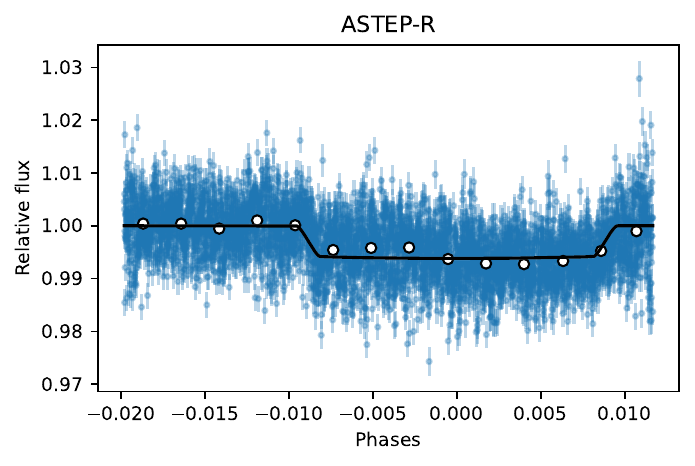}
\end{center}
\caption{Ground-based observation of transit of TIC\,147027702 (\textit{1$^{\rm st}$ panel}) and TIC\,245076932 in $B$ filter (\textit{2$^{\rm nd}$ panel}) and $R$ filter (\textit{3$^{\rm rd}$ panel}) with the best model (black). White points show the binned light curve.}
\label{fig:tic147_ground}
\end{figure*}

\begin{figure*}[h]
\begin{center}
\includegraphics[width=0.3\linewidth]{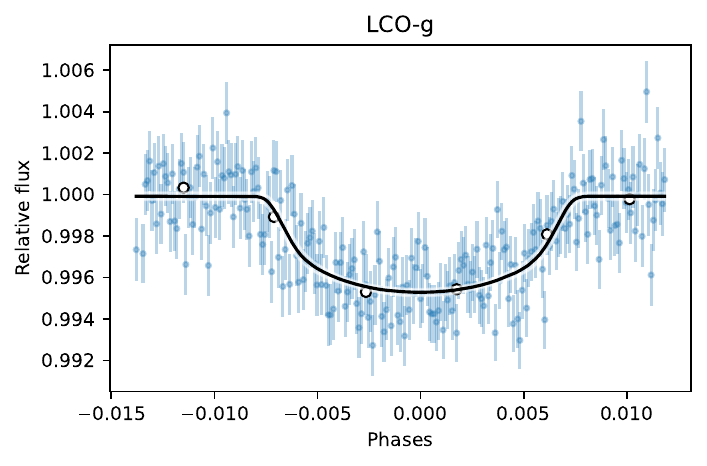}
\includegraphics[width=0.3\linewidth]{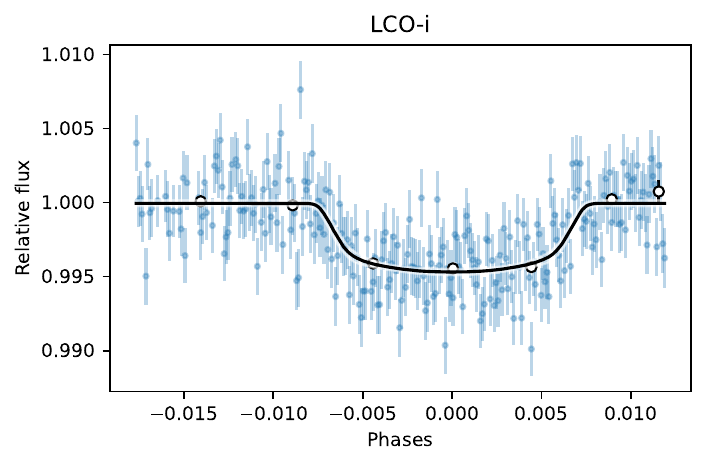}
\includegraphics[width=0.3\linewidth]{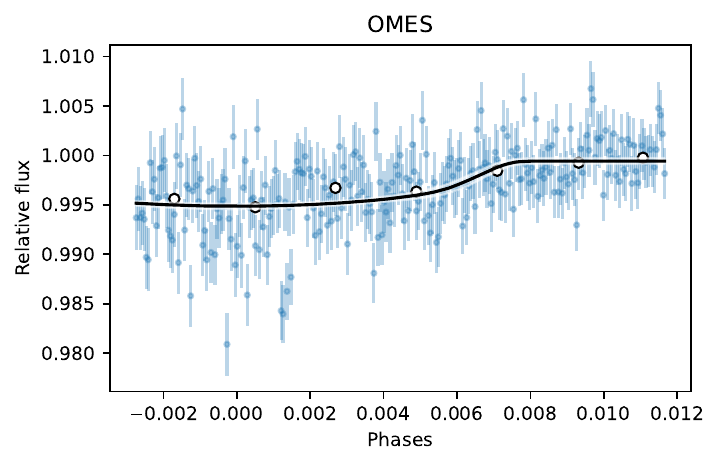}
\end{center}
\caption{Ground-based observations of transit of TIC\,87422071 with the best model (black). White points show the binned light curve. \textit{From left to right:} observation from LCO in $g'$ filter and $i'$ filter, observation from OMES in $r'$ filter.}
\label{fig:tic874_ground}
\end{figure*}

A transit of TIC\,147027702 was observed with OMES during the night of 4/5 April 2025 using a Sloan $r'$ filter and adopting exposure times of 16~s. The data were processed with an automated dedicated pipeline to deliver the relative photometric light curve of the event. This observation registered an ingress of the transit at the expected time. The light curve obtained, together with a theoretical one based on the model described in Sect.~\ref {model}, is shown in Fig.~\ref{fig:tic147_ground}.

A transit of TIC\,245076932 was monitored with the ASTEP telescope during the night of 4/5 August 2025 using Johnson $B$ and $R$ filters. The adopted exposure time was 60~s in the $B$ filter and 8~s in the $R$ one. The complete transit was captured (see Fig.~\ref{fig:tic147_ground}).

Two ground-based transits of TIC\,87422071 were observed (see Fig.~\ref{fig:tic874_ground}). 
We observed a full transit window on 27/28 May 2021 with the Sloan $g'$ and Sloan $i'$ bands from the LCOGT telescope. The images were calibrated using the standard LCOGT {\tt BANZAI} pipeline \citep{McCully:2018} and differential photometric data were extracted using {\tt AstroImageJ} \citep{Collins:2017}. We used circular $5\farcs8$ photometric apertures that excluded flux from the nearest Gaia DR3 catalogue neighbour that is bright enough to possibly be the source of the TESS transit detection (Gaia DR3 6721152024675106816). We detected the transit in the target star photometric aperture in both bands, which confirms that the TESS-detected event is indeed occurring in TIC\,87422071.
Another partial transit was captured in the Observatoire Moana during the night of 26/27 March 2024 using a Sloan $r'$ filter and exposure times of 60 s. Photometric data were processed with the same pipeline as in the case of TIC\,147027702.

In all three cases, we detected the transit exactly on the target studied. The shape and time of observed transit agree with the theoretical model as is clearly visible in Fig.~\ref{fig:tic147_ground} and \ref{fig:tic874_ground}.

\subsection{Spectroscopy}

We obtained follow-up radial velocities (RV) using the new PLATOSpec spectrograph \citep{Kabath2025}. PLATOSpec is a white pupil echelle fibre-fed spectrograph with a spectral resolution power of $R=70,000$ covering a wavelength range of 380 to 700 nm. PLATOSpec is attached to the ESO 1.52~m telescope at the La Silla Observatory, in Chile. Currently, the use of simultaneous calibrations using a Thorium-Argon (ThAr) lamp in a second fibre allows us to obtain RVs with a precision as good as 6~m/s for bright solar-type stars with slow projected rotational velocities. PLATOSpec data were processed with a modified version of the \texttt{ceres} pipeline \citep{ceres}, which performs all the reduction and processing steps required to obtain precision RVs starting from raw science and calibration images. RV measurements were calculated using cross-correlation with a G2-type star template. Besides precision RVs, the pipeline also delivers other outputs associated with the cross-correlation peak, like bisector span (BIS) and full-width at half maximum (FWHM) measurements. 

\begin{figure}
    \centering
    \includegraphics[width=\columnwidth]{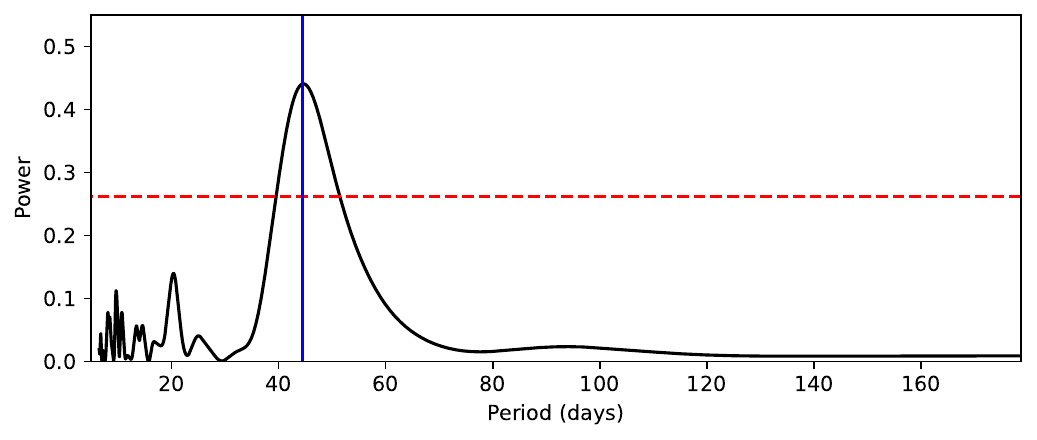}
    \includegraphics[width=\columnwidth]{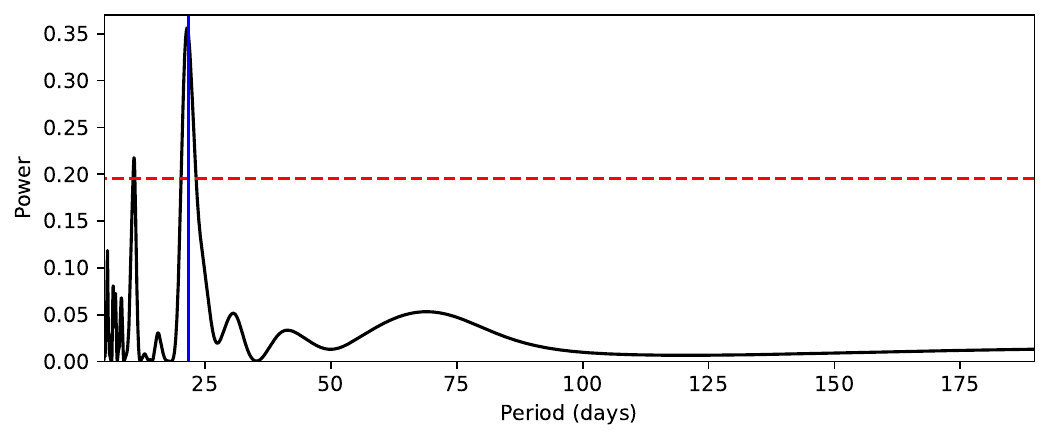}
    \includegraphics[width=\columnwidth]{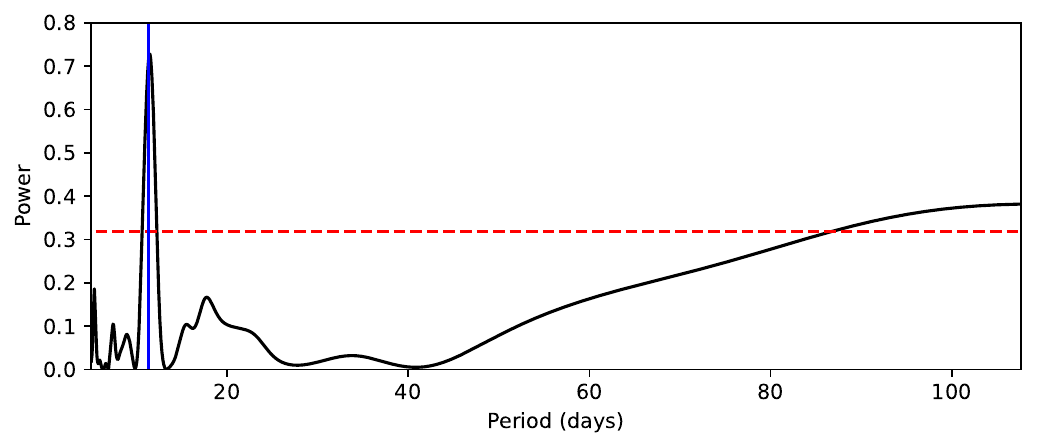}
    \caption{Periodograms of RV data of TIC\,147027702 (\textit{top}), TIC\,245076932 (\textit{middle}) and TIC\,87422071 (\textit{bottom}). The horizontal red line corresponds to the 1\% false alarm probability, and the blue line marks the orbital period determined from the photometric data.}
    \label{fig:gls}
\end{figure}

We obtained 55 usable spectra of TIC\,147027702 between January and July 2025 with a mean RV uncertainty of 26.6~m/s (listed in Tab.~\ref{tab:tic147_rv}), which was due to the moderate faintness and projected rotational velocity of the star. The observations were performed with exposure times of 20 and 30 minutes, depending on weather conditions. The average signal-to-noise ratio (SNR) per resolution element of these spectra was 42. A Generalized Lomb-Scargle (GLS) periodogram (Fig.~\ref{fig:gls}) of the radial velocities shows a significant peak at $\approx44.6$ days, which is consistent with the orbital period of the candidate found from the TESS light curves. Additionally, we found no significant correlation between RVs and BIS (see Fig.~\ref{fig:bis}). RV measurements do not correlate with values of FWHM, CCF width or depth, or SNR (correlation coefficient less than 0.05).

We observed TIC\,245076932 between February and August 2025 and collected 79 spectra suitable for the following analysis (listed in Tab.~\ref{tab:tic245_rv}). Observations performed with exposure times of 30 or 40 minutes resulted in a mean SNR of 34 and a mean RV uncertainty of 16.6~m/s. A significant peak at 21.4 days was detected in a GLS periodogram of RVs (Fig.~\ref{fig:gls}), which is in good agreement with the orbital period detected by TESS photometry. An additional strong peak is visible at around 10.9 days, which is close to half of the main period. We found no significant correlation between RVs and BIS, FWHM, shape of CCF or SNR (correlation coefficient less than 0.1; see the relation for BIS in Fig.~\ref{fig:bis}).

For TIC\,87422071, we obtained 43 spectra between May and August 2025 using PLATOSpec (listed in Tab.~\ref{tab:tic874_rv}). An exposure time of 40 minutes was used. The mean value of the RV uncertainty is 23.0~m/s, and the mean SNR is 33.
Additionally, four spectra obtained by the FEROS spectrograph \citep[Fiber-fed Extended Range Optical Spectrograph;][]{feros} in July 2024 and June 2025 with an exposure time of 20 minutes were used (see Tab.~\ref{tab:tic874_rv-feros}). FEROS is a spectrograph that covers the range of 360 to 920 nm with a mean resolving power of $R=48,000$ installed in the ESO-MPG 2.2 telescope at La Silla. The mean SNR is 48 and the RV uncertainty is 17.2~m/s. 
Using the GLS periodogram of the RVs (Fig.~\ref{fig:gls}), we detected a significant peak at 11.5 days very close to the photometric orbital period (11.36 days). A possible weak correlation between RVs and BIS could be observed (Fig.~\ref{fig:bis}; with a correlation coefficient of 0.4). This correlation is driven by two specific measurements obtained under poor observing conditions.

\section{Stellar parameters}
\label{star}
We followed the same iterative procedure adopted in \cite{brahm:2023} to compute the atmospheric and physical parameters of the host star. We generated a co-added high SNR PLATOSpec spectrum of individual targets. We used the \texttt{zaspe} code \citep{zaspe} to obtain a first estimation of the atmospheric parameters ($T_{\rm eff}$, $\log g$, $[$Fe/H$]$, and $v\sin i$). Then we performed a spectral energy distribution (SED) fit to publicly available broadband photometry using the PARSEC stellar evolution models \citep{parsec} and the GAIA DR3 parallax \citep{gaia}. In this step, we adopt the metallicity value found in our spectroscopic analysis. The SED fit delivers a more precise value for $\log g$, so we proceed with a new \texttt{zaspe} run where $\log g$ is fixed to the value obtained with the SED fit. We keep iterating between these two procedures until we reach convergence in the derived atmospheric parameters of two contiguous \texttt{zaspe} runs. To derive realistic uncertainties for the determined parameters, we adopted the procedure described in the paper of \cite{Tayar2022} and accordingly inflated the formal uncertainties obtained from our spectral modelling.

According to this modelling, TIC\,147027702 is a slightly metal poor F-type dwarf star ($[$Fe/H$]=-0.08 \pm 0.05$ dex; $T_{\rm eff} = 6410 \pm 183$\,K) with a moderate projected rotational velocity of $v\sin i = 8.8 \pm 0.3$,km/s. TIC\,245076932 is a late F-type dwarf star ($T_{\rm eff} = 6130 \pm 178$\,K) with a rotational velocity of $v\sin i = 4.2 \pm 0.3$ km/s close to the end of its life as a main-sequence star (age of about 4.5 Gyr). TIC\,87422071 is a slightly metal rich dwarf star of the F-type ($[$Fe/H$] = 0.16 \pm 0.05$ dex; $T_{\rm eff} = 6150 \pm 178$\,K), very similar to TIC\,245076932 but 1 Gyr younger. Its rotational velocity is $v\sin i = 7.1 \pm 0.3$ km/s.

Table~\ref{tab:star} shows the main stellar parameters of the targets and their identifiers in various catalogues. The sky coordinates (RA, Dec), the proper motion in both directions ($\mu_\alpha$, $\mu_\delta$) and the parallax $\pi$ were obtained from Gaia DR3. All listed magnitudes (except for the \textit{G} one) were collected from the TESS Input Catalog \citep[TIC;][]{tic}.

\begin{table*}[h]
\caption{Stellar parameters and identifiers.}
\label{tab:star}
\begin{center}
\begin{tabular}{lcccc}
	\hline\hline
	                                   &      TIC\,147027702       &     TIC\,245076932     &       TIC\,87422071       & Source \\ \hline
	TIC ID                             &         147027702         &       245076932        &         87422071          &   1    \\
	TOI                                &            ---            &          ---           &           6752            &   1    \\
	TYC                                &        7735-1496-1        &      8259-1806-1       &        7912-829-1         &   2    \\
	2MASS                              &     J10470550-4355545     &   J13121527-5112519    &     J18120831-4414359     &   2    \\
	Gaia DR3                           &    5390877409019489408    &  6080787214190400384   &    6721152093391972480    &   2    \\ \hline
	RA (J2000.0)                       &       10h47m05.50s        &      13h12m15.28s      &       18h12m08.32s        &   3    \\
	Dec (J2000.0)                      &      $-$43d55m54.58s      &    $-$51d12m51.99s     &      $-$44d14m35.95s      &   3    \\
	$\mu_\alpha$ (mas/yr)              &    $-11.805 \pm 0.013$    &  $-24.412 \pm 0.021$   &    $-9.263 \pm 0.018$     &   3    \\
	$\mu_\delta$ (mas/yr)              &     $7.228 \pm 0.015$     &   $ 2.204 \pm 0.017$   &    $-19.588 \pm 0.014$    &   3    \\
	$\pi$ (mas)                        &    $2.1893 \pm 0.0170$    &  $3.1788 \pm 0.0226$   &    $2.5057 \pm 0.0209$    &   3    \\
	Distance (pc)                      &     $456.8 \pm 3.5 $      &    $314.7 \pm  2.2$    &      $399.1 \pm 3.3$      &   5    \\ \hline
	T (mag)                            &    $10.630 \pm 0.006$     &   $11.036 \pm 0.006$   &    $11.194 \pm  0.006$    &   4    \\
	B (mag)                            &    $11.570 \pm 0.122$     &   $12.203 \pm 0.251$   &    $12.565 \pm 0.322$     &   4    \\
	V (mag)                            &    $11.286 \pm 0.009$     &   $11.727 \pm 0.017$   &    $11.982 \pm 0.022$     &   4    \\
	G (mag)                            &    $10.990 \pm 0.003$     &   $11.463 \pm 0.003$   &    $11.607 \pm 0.003$     &   3    \\
	J (mag)                            &    $10.110 \pm 0.024$     &   $10.424 \pm 0.024$   &    $10.588 \pm 0.022$     &   4    \\
	H (mag)                            &     $9.868 \pm 0.026$     &   $10.155 \pm 0.022$   &    $10.376 \pm 0.023$     &   4    \\
	K (mag)                            &     $9.801 \pm 0.019$     &   $10.083 \pm 0.019$   &    $10.223 \pm 0.019$     &   4    \\ \hline
	Luminosity (L$_\sun$)              &  $6.80^{+0.52}_{-0.42}$   &    $2.11 \pm 0.13$     &  $2.80^{+0.26}_{-0.21}$   &   5    \\
	Mass (M$_\sun$)                    & $1.406^{+0.074}_{-0.086}$ &   $1.117 \pm 0.063$    & $1.239^{+0.070}_{-0.080}$ &   5    \\
	Radius (R$_\sun$)                  &     $2.136 \pm 0.096$     &   $1.288 \pm 0.058$    & $1.489^{+0.072}_{-0.071}$ &   5    \\
	$\rho_\star$ (kg/m$^3$)            &   $201.6^{+3.7}_{-2.2}$   & $736.7^{+1.8}_{-1.3}$  &   $530.6^{+2.1}_{-1.8}$   &   5    \\
	$T_{\rm eff}$ (K)                  &      $6410 \pm 183$       &     $6130 \pm 178$     &      $6150 \pm 178$       &   5    \\
	Age (Gyr)                          &    $2.7^{+0.7}_{-0.6}$    &  $4.5^{+1.3}_{-1.2}$   &    $3.5^{+1.3}_{-0.9}$    &   5    \\
	$\log g$ (cm/s$^2$)                &     $3.926 \pm 0.020$     &   $4.267 \pm 0.020$    & $4.183^{+0.022}_{-0.023}$ &   5    \\
	$A_v$ (mag)                        &  $0.20^{+0.09}_{-0.08}$   & $0.30^{+0.07}_{-0.08}$ &      $0.25 \pm 0.11$      &   5    \\
	$[$Fe/H$]$ (dex)                   &     $-0.08 \pm 0.05$      &    $0.02 \pm 0.05$     &      $0.16 \pm 0.05$      &   5    \\
	$v\sin i$ (km/s)                   &       $8.8 \pm 0.3$       &     $4.2 \pm 0.3$      &       $7.1 \pm 0.3$       &   5    \\
	$P_{\rm rot}$ (upper limit) (days) &      $12.3 \pm 0.6$       &     $15.5 \pm 1.4$     &      $10.6 \pm 0.7$       &   5    \\ \hline
\end{tabular}
\end{center}
\textbf{References:} (1) TESS \citep{tess}, (2) SIMBAD Database \citep{simbad}, (3) Gaia DR3 \citep{gaia}, (4) TIC v8.2 \citep{tic}, (5) this work.
\end{table*}

\section{Analysis and results}
\label{model}

We performed a joint analysis of all photometric data (TESS and ground-based transits) and RV measurements using the \texttt{juliet} package \citep{juliet} based on the packages \texttt{radvel} \citep[RV modelling;][]{radvel} and \texttt{batman} \citep[transit modelling;][]{batman}.

\begin{table}[h]
\caption{Prior parameter distributions and median value of the posterior distributions of the fitted parameters of the joint transit and RV analysis of TIC\,147027702.}
\label{tab:tic147_prior}
\begin{tabular}{lcc}
	\hline\hline
	Parameter                      &          Distribution           &               Value                \\ \hline
	$P$ (days)                     &   $\mathcal{N}$ (44.405, 0.2)   & $44.405236^{+1.33e-4}_{-1.27e-4}$  \\
	$t_0 - 2457000$ (days)         &   $\mathcal{N}$ (1550.7, 0.2)   & $1550.73135^{+3.47e-3}_{-3.82e-3}$ \\
	$b \equiv a/R_\star \cos i$    &      $\mathcal{U}$ (0, 1)       &     $0.230^{+0.145}_{-0.155}$      \\
	$p \equiv R_p / R_\star$       &   $\mathcal{U}$ (0.001, 0.1)    &    $0.0469^{+0.0009}_{-0.0010}$    \\
	$e\sin\omega$                  &   $\mathcal{U}$ ($-$0.8, 0.8)   &     $-0.076^{+0.026}_{-0.048}$     \\
	$e\cos\omega$                  &   $\mathcal{U}$ ($-$0.8, 0.8)   &     $-0.087^{+0.050}_{-0.050}$     \\
	$\rho_\star$ (kg/m$^3$)        &     $\mathcal{N}$ (200, 15)     &      $201.55^{+3.67}_{-2.17}$      \\
	$K$ (m/s)                      &     $\mathcal{U}$ (0, 100)      &      $50.43^{+1.36}_{-3.61}$       \\
	$\mu_{\rm PlatoSpec}$ (m/s)    &  $\mathcal{U}$ (22500, 23000)   &     $22753.56^{+1.30}_{-2.68}$     \\
	$\sigma_{\rm PlatoSpec}$ (m/s) &     $\mathcal{J}$ (0, 1000)     &      $16.39^{+5.72}_{-1.70}$       \\
	$q_{1, \rm TESS}$              &      $\mathcal{U}$ (0, 1)       &       $0.14^{+0.18}_{-0.08}$       \\
	$q_{2, \rm TESS}$              &      $\mathcal{U}$ (0, 1)       &       $0.42^{+0.36}_{-0.29}$       \\
	$q_{1, \rm ground-r}$          &      $\mathcal{U}$ (0, 1)       &       $0.66^{+0.24}_{-0.39}$       \\
	$q_{2, \rm ground-r}$          &      $\mathcal{U}$ (0, 1)       &       $0.33^{+0.35}_{-0.24}$       \\
	$\mu_{\rm TESS-S9}$            &     $\mathcal{N}$ (0, 0.1)      &  $-1.04e-4^{+9.28e-4}_{-5.86e-4}$  \\
	$\mu_{\rm TESS-S10}$           &     $\mathcal{N}$ (0, 0.1)      & $-2.83e-4^{+11.72e-4}_{-10.92e-4}$ \\
	$\mu_{\rm TESS-S63}$           &     $\mathcal{N}$ (0, 0.1)      &  $-1.08e-4^{+0.66e-4}_{-0.49e-4}$  \\
	$\mu_{\rm TESS-S90}$           &     $\mathcal{N}$ (0, 0.1)      &  $-2.00e-4^{+6.87e-4}_{-7.57e-4}$  \\
	$\mu_{\rm ground-r}$           &     $\mathcal{N}$ (0, 0.1)      & $-13.75e-4^{+1.29e-4}_{-1.42e-4}$  \\
	$\sigma_{\rm TESS-S9}$         &    $\mathcal{J}$ (0.1, 1000)    &       $1.89^{+0.79}_{-1.58}$       \\
	$\sigma_{\rm TESS-S10}$        &    $\mathcal{J}$ (0.1, 1000)    &      $218.29^{+2.72}_{-2.25}$      \\
	$\sigma_{\rm TESS-S63}$        &    $\mathcal{J}$ (0.1, 1000)    &       $3.46^{+3.44}_{-2.04}$       \\
	$\sigma_{\rm TESS-S90}$        &    $\mathcal{J}$ (0.1, 1000)    &       $2.87^{+1.58}_{-2.19}$       \\
	$\sigma_{\rm ground-r}$        &    $\mathcal{J}$ (0.1, 1000)    &      $989.77^{+2.35}_{-0.83}$      \\
	$\sigma_{\rm GP, TESS-S9}$     & $\mathcal{J}$ ($10^{-8}$, 1000) &  $6.90e-4^{+14.51e-4}_{-3.19e-4}$  \\
	$\sigma_{\rm GP, TESS-S10}$    & $\mathcal{J}$ ($10^{-8}$, 1000) &  $9.10e-4^{+18.37e-4}_{-5.34e-4}$  \\
	$\sigma_{\rm GP, TESS-S63}$    & $\mathcal{J}$ ($10^{-8}$, 1000) &  $0.18e-4^{+2.74e-4}_{-0.17e-4}$   \\
	$\sigma_{\rm GP, TESS-S90}$    & $\mathcal{J}$ ($10^{-8}$, 1000) &  $9.33e-4^{+0.49e-4}_{-1.14e-4}$   \\
	$\rho_{\rm GP, TESS-S9}$       &   $\mathcal{J}$ (0.001, 1000)   &       $1.66^{+1.63}_{-0.85}$       \\
	$\rho_{\rm GP, TESS-S10}$      &   $\mathcal{J}$ (0.001, 1000)   &       $2.76^{+2.66}_{-1.33}$       \\
	$\rho_{\rm GP, TESS-S63}$      &   $\mathcal{J}$ (0.001, 1000)   &       $1.08^{+0.37}_{-0.94}$       \\
	$\rho_{\rm GP, TESS-S90}$      &   $\mathcal{J}$ (0.001, 1000)   &       $1.13^{+0.38}_{-0.35}$       \\ \hline
\end{tabular}
\end{table}

\begin{table}[h]
\caption{Prior parameter distributions for TIC\,245076932. For detail description see Tab.~\ref{tab:tic147_prior}.}
\label{tab:tic245_prior}
\begin{tabular}{lcc}
	\hline\hline
	Parameter                      &          Distribution           &               Value                \\ \hline
	$P$ (days)                     &   $\mathcal{N}$ (21.614, 0.2)   &       $21.613890 \pm 1.7e-5$       \\
	$t_0 - 2457000$ (days)         &   $\mathcal{N}$ (2357.6, 0.2)   & $2357.66168^{+1.19e-3}_{-1.26e-3}$ \\
	$b \equiv a/R_\star \cos i$    &      $\mathcal{U}$ (0, 1)       &     $0.106^{+0.106}_{-0.068}$      \\
	$p \equiv R_p / R_\star$       &   $\mathcal{U}$ (0.001, 0.1)    &        $0.0774 \pm 0.0004$         \\
	$e\sin\omega$                  &   $\mathcal{U}$ ($-$0.8, 0.8)   &     $-0.386^{+0.008}_{-0.016}$     \\
	$e\cos\omega$                  &   $\mathcal{U}$ ($-$0.8, 0.8)   &     $0.182^{+0.026}_{-0.030}$      \\
	$\rho_\star$ (kg/m$^3$)        &     $\mathcal{N}$ (740, 60)     &      $736.66^{+1.82}_{-1.32}$      \\
	$K$ (m/s)                      &     $\mathcal{U}$ (0, 100)      &      $38.53^{+1.51}_{-1.93}$       \\
	$\mu_{\rm PlatoSpec}$ (m/s)    &  $\mathcal{U}$ (-6000, -3000)   &     $-4169.24^{+2.87}_{-2.17}$     \\
	$\sigma_{\rm PlatoSpec}$ (m/s) &     $\mathcal{J}$ (0, 1000)     &      $20.57^{+1.13}_{-1.66}$       \\
	$q_{1, \rm TESS}$              &      $\mathcal{U}$ (0, 1)       &       $0.24^{+0.15}_{-0.10}$       \\
	$q_{2, \rm TESS}$              &      $\mathcal{U}$ (0, 1)       &       $0.23^{+0.28}_{-0.15}$       \\
	$q_{1, \rm ground-B}$          &      $\mathcal{U}$ (0, 1)       &       $0.20^{+0.04}_{-0.03}$       \\
	$q_{2, \rm ground-B}$          &      $\mathcal{U}$ (0, 1)       &       $0.87^{+0.09}_{-0.12}$       \\
	$q_{1, \rm ground-R}$          &      $\mathcal{U}$ (0, 1)       &     $0.008^{+0.012}_{-0.005}$      \\
	$q_{2, \rm ground-R}$          &      $\mathcal{U}$ (0, 1)       &       $0.61^{+0.30}_{-0.35}$       \\
	$\mu_{\rm TESS-S11}$           &     $\mathcal{N}$ (0, 0.1)      &  $-9.33e-4^{+14.9e-4}_{-5.96e-4}$  \\
	$\mu_{\rm TESS-S37}$           &     $\mathcal{N}$ (0, 0.1)      &     $0.182^{+0.066}_{-0.058}$      \\
	$\mu_{\rm TESS-S38}$           &     $\mathcal{N}$ (0, 0.1)      & $-4.11e-4^{+90.19e-4}_{-6.75e-4}$  \\
	$\mu_{\rm TESS-S64}$           &     $\mathcal{N}$ (0, 0.1)      &     $0.149^{+0.057}_{-0.048}$      \\
	$\mu_{\rm ground-B}$           &     $\mathcal{N}$ (0, 0.1)      &  $-6.17e-4^{+0.35e-4}_{-0.32e-4}$  \\
	$\mu_{\rm ground-R}$           &     $\mathcal{N}$ (0, 0.1)      &  $-0.19e-4^{+0.63e-4}_{-0.61e-4}$  \\
	$\sigma_{\rm TESS-S11}$        &    $\mathcal{J}$ (0.1, 1000)    &       $5.18^{2.88}_{-4.04}$        \\
	$\sigma_{\rm TESS-S37}$        &    $\mathcal{J}$ (0.1, 1000)    &       $4.28^{+2.82}_{-3.87}$       \\
	$\sigma_{\rm TESS-S38}$        &    $\mathcal{J}$ (0.1, 1000)    &       $4.86^{+2.54}_{-3.40}$       \\
	$\sigma_{\rm TESS-S64}$        &    $\mathcal{J}$ (0.1, 1000)    &       $5.10^{+4.25}_{-4.07}$       \\
	$\sigma_{\rm ground-B}$        &    $\mathcal{J}$ (0.1, 1000)    &      $15.18^{+2.03}_{-0.55}$       \\
	$\sigma_{\rm ground-R}$        &    $\mathcal{J}$ (0.1, 1000)    &      $14.93^{+2.13}_{-1.89}$       \\
	$\sigma_{\rm GP, TESS-S11}$    & $\mathcal{J}$ ($10^{-8}$, 1000) &  $9.78e-4^{+39.13e-4}_{-4.33e-4}$  \\
	$\sigma_{\rm GP, TESS-S37}$    & $\mathcal{J}$ ($10^{-8}$, 1000) &     $0.108^{+0.0987}_{-0.039}$     \\
	$\sigma_{\rm GP, TESS-S38}$    & $\mathcal{J}$ ($10^{-8}$, 1000) &  $16.25e-4^{+0.0134}_{-10.93e-4}$  \\
	$\sigma_{\rm GP, TESS-S64}$    & $\mathcal{J}$ ($10^{-8}$, 1000) &     $0.093^{+0.082}_{-0.035}$      \\
	$\rho_{\rm GP, TESS-S11}$      &   $\mathcal{J}$ (0.001, 1000)   &       $0.71^{+1.67}_{-0.46}$       \\
	$\rho_{\rm GP, TESS-S37}$      &   $\mathcal{J}$ (0.001, 1000)   &      $34.95^{+2.56}_{-2.97}$       \\
	$\rho_{\rm GP, TESS-S38}$      &   $\mathcal{J}$ (0.001, 1000)   &       $1.37^{+2.90}_{-1.03}$       \\
	$\rho_{\rm GP, TESS-S64}$      &   $\mathcal{J}$ (0.001, 1000)   &      $39.17^{+3.55}_{-2.79}$       \\ \hline
\end{tabular}
\end{table}

\begin{table}[h]
\caption{Prior parameter distributions for TIC\,87422071. For detail description see Tab.~\ref{tab:tic147_prior}.}
\label{tab:tic874_prior}
\begin{tabular}{lcc}
	\hline\hline
	Parameter                      &          Distribution           &               Value                \\ \hline
	$P$ (days)                     &   $\mathcal{N}$ (11.365, 0.2)   & $11.364929^{+0.21e-4}_{-0.22e-4}$  \\
	$t_0 - 2457000$ (days)         &   $\mathcal{N}$ (1680.7, 0.2)   & $1680.65985^{+2.58e-3}_{-2.84e-3}$ \\
	$b \equiv a/R_\star \cos i$    &      $\mathcal{U}$ (0, 1)       &     $0.718^{+0.073}_{-0.102}$      \\
	$p \equiv R_p / R_\star$       &   $\mathcal{U}$ (0.001, 0.1)    &    $0.0669^{+0.0027}_{-0.0027}$    \\
	$e\sin\omega$                  &   $\mathcal{U}$ ($-$0.8, 0.8)   &     $0.056^{+0.110}_{-0.114}$      \\
	$e\cos\omega$                  &   $\mathcal{U}$ ($-$0.8, 0.8)   &     $0.046^{+0.067}_{-0.071}$      \\
	$\rho_\star$ (kg/m$^3$)        &     $\mathcal{N}$ (530, 50)     &      $530.57^{+2.11}_{-1.80}$      \\
	$K$ (m/s)                      &     $\mathcal{U}$ (0, 200)      &      $101.63^{+3.13}_{-3.40}$      \\
	$\mu_{\rm PlatoSpec}$ (m/s)    &   $\mathcal{U}$ (1000, 3000)    &     $2759.65^{+2.14}_{-2.76}$      \\
	$\sigma_{\rm PlatoSpec}$ (m/s) &     $\mathcal{J}$ (0, 1000)     &      $44.95^{+1.62}_{-1.95}$       \\
	$\mu_{\rm FEROS}$ (m/s)        &   $\mathcal{U}$ (1000, 3000)    &     $2735.61^{+3.28}_{-3.11}$      \\
	$\sigma_{\rm FEROS}$ (m/s)     &     $\mathcal{J}$ (0, 1000)     &      $34.94^{+3.09}_{-2.50}$       \\
	$q_{1, \rm TESS}$              &      $\mathcal{U}$ (0, 1)       &       $0.22^{+0.27}_{-0.15}$       \\
	$q_{2, \rm TESS}$              &      $\mathcal{U}$ (0, 1)       &       $0.47^{+0.40}_{-0.33}$       \\
	$q_{1, \rm ground-g}$          &      $\mathcal{U}$ (0, 1)       &       $0.65^{+0.22}_{-0.30}$       \\
	$q_{2, \rm ground-g}$          &      $\mathcal{U}$ (0, 1)       &       $0.51^{+0.29}_{-0.35}$       \\
	$q_{1, \rm ground-r}$          &      $\mathcal{U}$ (0, 1)       &       $0.63^{+0.26}_{-0.33}$       \\
	$q_{2, \rm ground-r}$          &      $\mathcal{U}$ (0, 1)       &       $0.66^{+0.24}_{-0.39}$       \\
	$q_{1, \rm ground-i}$          &      $\mathcal{U}$ (0, 1)       &       $0.43^{+0.39}_{-0.29}$       \\
	$q_{2, \rm ground-i}$          &      $\mathcal{U}$ (0, 1)       &       $0.35^{+0.37}_{-0.26}$       \\
	$\mu_{\rm TESS-S13}$           &     $\mathcal{N}$ (0, 0.1)      &    $0.0015^{+0.0027}_{-0.0025}$    \\
	$\mu_{\rm TESS-S66}$           &     $\mathcal{N}$ (0, 0.1)      &    $0.3361^{+0.0846}_{-0.0865}$    \\
	$\mu_{\rm TESS-S93}$           &     $\mathcal{N}$ (0, 0.1)      & $-0.64e-4^{+2.14e-4}_{-11.38e-4}$  \\
	$\mu_{\rm ground-g}$           &     $\mathcal{N}$ (0, 0.1)      &  $0.78e-4^{+1.82e-4}_{-1.80e-4}$   \\
	$\mu_{\rm ground-r}$           &     $\mathcal{N}$ (0, 0.1)      &  $5.64e-4^{+3.06e-4}_{-3.09e-4}$   \\
	$\mu_{\rm ground-i}$           &     $\mathcal{N}$ (0, 0.1)      &  $0.59e-4^{+1.92e-4}_{-2.01e-4}$   \\
	$\sigma_{\rm TESS-S13}$        &    $\mathcal{J}$ (0.1, 1000)    &       $9.52^{+2.13}_{-2.68}$       \\
	$\sigma_{\rm TESS-S66}$        &    $\mathcal{J}$ (0.1, 1000)    &      $997.07^{+1.89}_{-1.97}$      \\
	$\sigma_{\rm TESS-S93}$        &    $\mathcal{J}$ (0.1, 1000)    &       $4.21^{+1.92}_{-3.29}$       \\
	$\sigma_{\rm ground-g}$        &    $\mathcal{J}$ (0.1, 1000)    &      $40.57^{+1.95}_{-2.29}$       \\
	$\sigma_{\rm ground-r}$        &    $\mathcal{J}$ (0.1, 1000)    &      $40.65^{+3.03}_{-3.12}$       \\
	$\sigma_{\rm ground-i}$        &    $\mathcal{J}$ (0.1, 1000)    &      $40.73^{+2.42}_{-2.28}$       \\
	$\sigma_{\rm GP, TESS-S13}$    & $\mathcal{J}$ ($10^{-8}$, 1000) &    $0.0038^{+0.0034}_{-0.0012}$    \\
	$\sigma_{\rm GP, TESS-S66}$    & $\mathcal{J}$ ($10^{-8}$, 1000) &    $0.1749^{+0.1732}_{-0.0665}$    \\
	$\sigma_{\rm GP, TESS-S93}$    & $\mathcal{J}$ ($10^{-8}$, 1000) &  $1.07e-4^{+61.13e-4}_{-1.02e-4}$  \\
	$\rho_{\rm GP, TESS-S13}$      &   $\mathcal{J}$ (0.001, 1000)   &       $1.35^{+0.85}_{-0.43}$       \\
	$\rho_{\rm GP, TESS-S66}$      &   $\mathcal{J}$ (0.001, 1000)   &      $537.36^{+3.74}_{-3.38}$      \\
	$\rho_{\rm GP, TESS-S93}$      &   $\mathcal{J}$ (0.001, 1000)   &      $11.51^{+2.38}_{-1.79}$       \\ \hline
\end{tabular}
\end{table}

The following set of planetary orbital parameters was used during the fitting routine as free ones: orbital period $P$, transit time $t_0$, impact parameter $b \equiv a/R_\star \cos i$, radius ratio $p \equiv R_p / R_\star$, eccentricity $e$ and argument of periastron $\omega$ (in forms $e\sin\omega$ and $e\cos\omega$), stellar density $\rho_\star$, and RV semi-amplitude $K$.

Each individual dataset is described by instrumental parameters. The relative flux offset $\mu$ and the jitter value $\sigma$ were used in the case of photometric data. We also adopted a quadratic limb-darkening law with coefficients $q_1$ and $q_2$. For RV measurements, we fitted instrumental systematic RV value~$\mu$ and jitter~$\sigma$.

Moreover, we utilised Gaussian processes \citep[GP;][]{GP} to model stellar activity. We use this method only with TESS data. The GP are described by two parameters -- amplitude $\sigma_{\rm GP}$ and time scale $\rho_{\rm GP}$. We used a simple (approximate) Matern kernel implemented in the \texttt{celerite} package \citep{celerite}.

Tables~\ref{tab:tic147_prior}, \ref{tab:tic245_prior} and \ref{tab:tic874_prior} list the priors of all fitted parameters for individual planetary candidates together with the median values of the posterior distributions. Uncertainties refer to the $1\sigma$ credibility interval.

\begin{table*}[h]
\caption{Fitted and derived planetary parameters of all systems.}
\label{tab:planet}
\begin{center}
\begin{tabular}{cccc}
	\hline\hline
	                            &           TIC\,147027702           &           TIC\,245076932           &            TIC\,87422071            \\ \hline
	        $P$ (days)          &       $44.40523 \pm 0.00013$       &      $21.613890 \pm 0.000017$      & $11.364929^{+0.000021}_{-0.000022}$ \\
	       $t_0$ (days)         & $2458550.7314^{+0.0035}_{-0.0038}$ & $2459357.6617^{+0.0012}_{-0.0013}$ & $2458680.6599^{+0.0026}_{-0.0028}$  \\
	$b \equiv a/R_\star \cos i$ &     $0.230^{+0.145}_{-0.155}$      &     $0.106^{+0.106}_{-0.068}$      &      $0.718^{+0.073}_{-0.102}$      \\
	 $p \equiv R_p / R_\star$   &    $0.0469^{+0.0009}_{-0.0010}$    &        $0.0774 \pm 0.0004$         &    $0.0669^{+0.0027}_{-0.0027}$     \\
	            $e$             &     $0.127^{+0.055}_{-0.048}$      &     $0.428^{+0.022}_{-0.019}$      &      $0.124^{+0.073}_{-0.060}$      \\
	      $\omega$ (deg)        &      $223.7^{+22.5}_{-17.4}$       &       $295.0^{+2.6}_{-3.3}$        &       $46.4^{+51.9}_{-93.6}$        \\
	  $\rho_\star$ (kg/m$^3$)   &      $201.55^{+3.67}_{-2.17}$      &      $736.66^{+1.82}_{-1.32}$      &      $530.57^{+2.11}_{-1.80}$       \\
	         $K$ (m/s)          &      $50.43^{+1.36}_{-3.61}$       &      $38.53^{+1.51}_{-1.93}$       &      $101.63^{+3.13}_{-3.40}$       \\ \hline
	         $i$ (deg)          &      $89.52^{+0.32}_{-0.30}$       &      $89.77^{+0.15}_{-0.23}$       &       $87.32^{+0.38}_{-0.27}$       \\
	       $M_p$ (M$_J$)        &       $1.09^{+0.07}_{-0.13}$       &       $0.51^{+0.04}_{-0.05}$       &       $1.29^{+0.10}_{-0.11}$        \\
	         $a$ (au)           &     $0.274^{+0.014}_{-0.013}$      &         $0.157 \pm 0.007$          &          $0.106 \pm 0.005$          \\
	       $R_p$ (R$_J$)        &     $0.975^{+0.065}_{-0.064}$      &         $0.970 \pm 0.048$          &      $0.969^{+0.087}_{-0.084}$      \\
	    $\rho_p$ (kg/m$^3$)     &     $1566.6^{+488.2}_{-426.5}$     &      $746.6^{+48.7}_{-55.9}$       &     $1878.5^{+330.2}_{-284.21}$     \\
	      $g_p$ (m/s$^2$)       &           $29.9 \pm 6.7$           &        $14.2^{+2.9}_{-2.6}$        &        $35.6^{+10.4}_{-8.3}$        \\
	     $T_{\rm eq}$ (K)       &         $863^{+26}_{-27}$          &            $845 \pm 25$            &            $1110 \pm 33$            \\
	 $\delta \equiv p^2$ (\%)   &         $0.220 \pm 0.009$          &         $0.599 \pm 0.006$          &      $0.447^{+0.036}_{-0.037}$      \\
	     $T_{14}$ (hours)       &        $15.9^{+0.9}_{-1.8}$        &           $9.2 \pm 0.5$            &         $3.6^{+1.7}_{-0.8}$         \\
	     $T_{23}$ (hours)       &        $14.4^{+0.9}_{-1.8}$        &           $7.9 \pm 0.4$            &         $2.8^{+1.4}_{-0.9}$         \\ \hline
\end{tabular}
\end{center}
\end{table*}

\begin{figure*}[h]
\begin{center}
\includegraphics[width=0.24\linewidth]{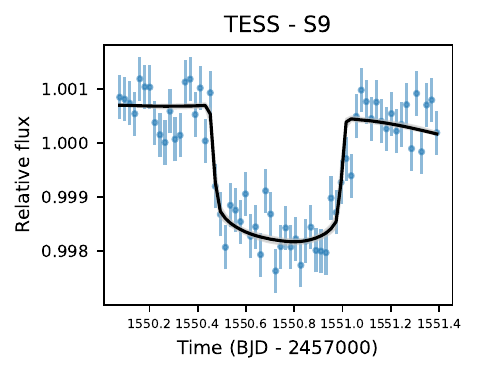}
\includegraphics[width=0.24\linewidth]{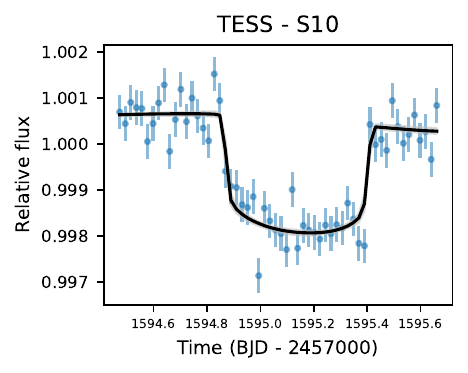}
\includegraphics[width=0.24\linewidth]{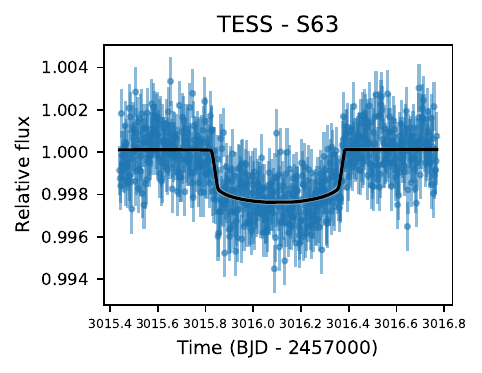}
\includegraphics[width=0.24\linewidth]{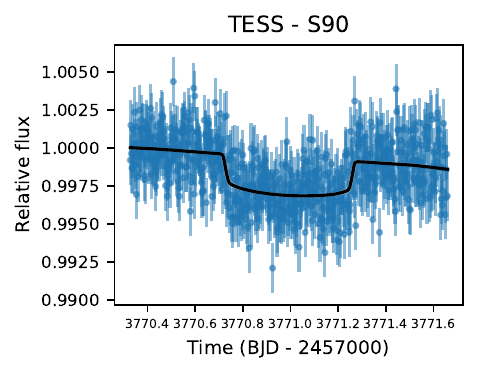}
\includegraphics[width=0.55\linewidth]{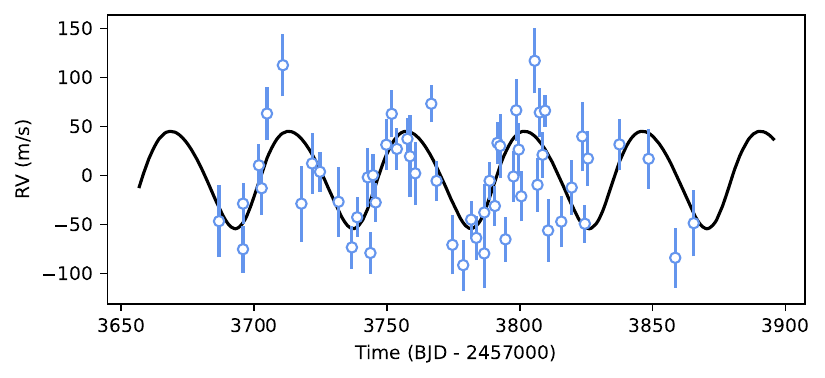}
\includegraphics[width=0.40\linewidth]{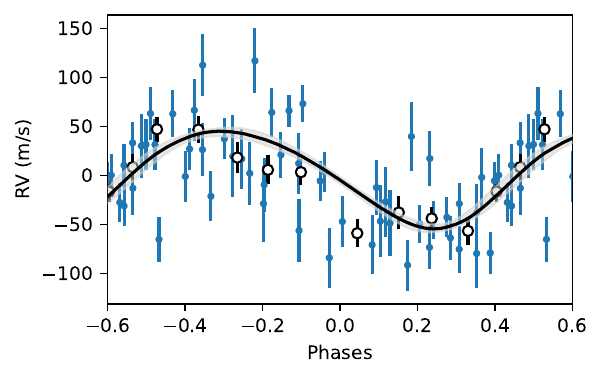}
\end{center}
\caption{\textit{Top:} Light curves of TIC\,147027702 for individual TESS sectors with the best model (black).
\textit{Bottom:} RV time series of TIC\,147027702 and phase-folded RV curve with the best model (black). White points show the binned RV curve.}
\label{fig:tic147}
\end{figure*}

\begin{figure*}[h]
\begin{center}
\includegraphics[width=0.19\linewidth]{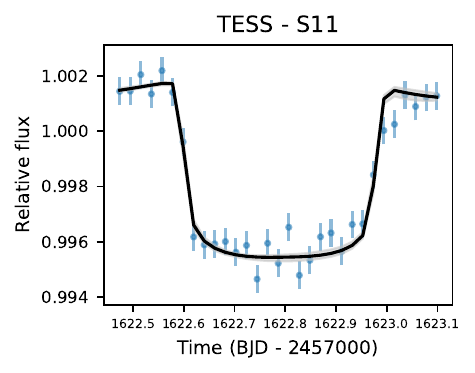}
\includegraphics[width=0.19\linewidth]{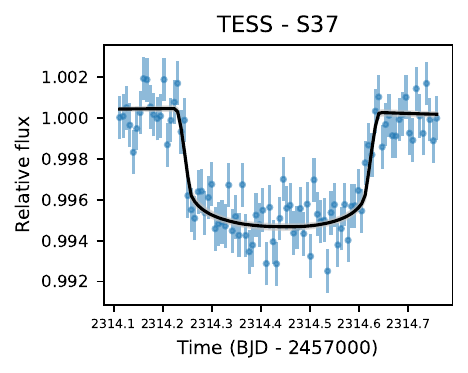}
\includegraphics[width=0.19\linewidth]{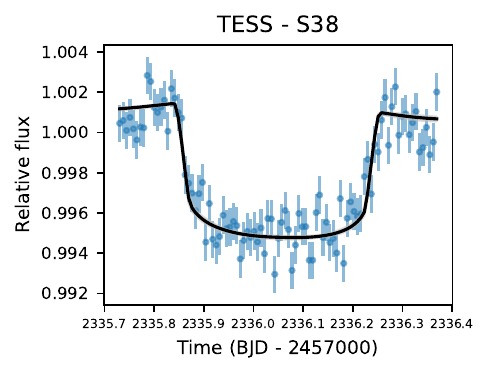}
\includegraphics[width=0.19\linewidth]{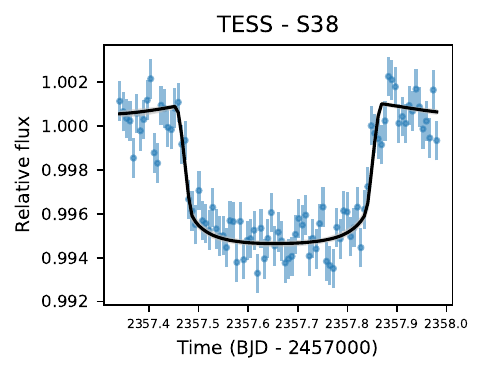}
\includegraphics[width=0.19\linewidth]{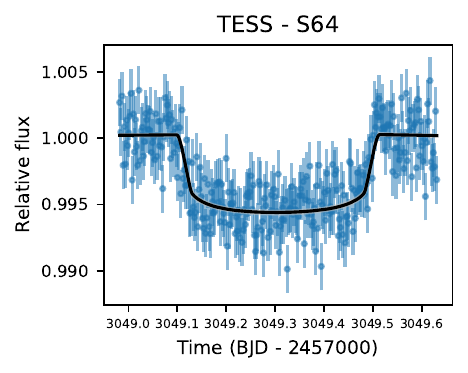}
\includegraphics[width=0.55\linewidth]{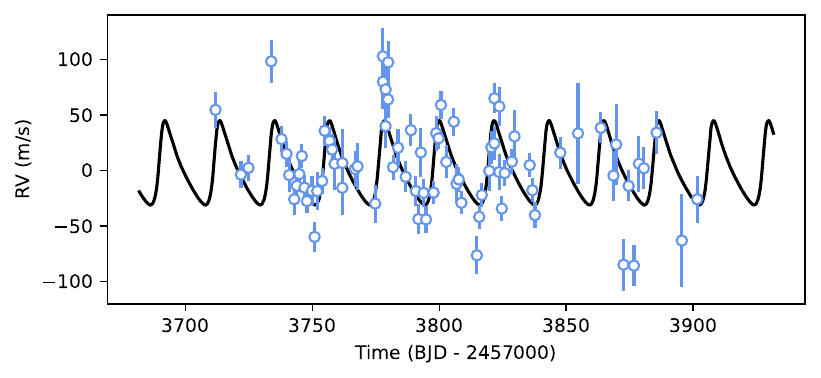}
\includegraphics[width=0.40\linewidth]{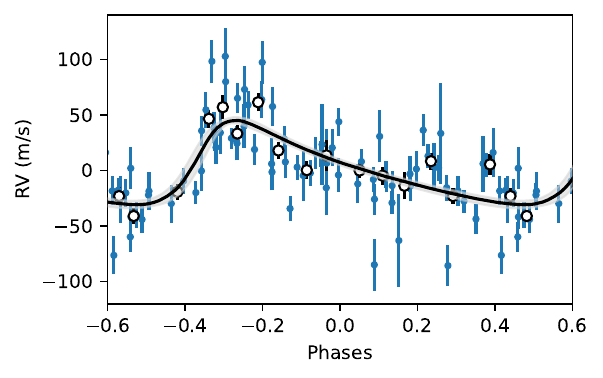}
\end{center}
\caption{Light curves and RV time series of TIC\,245076932. For detail description see Fig.~\ref{fig:tic147}.}
\label{fig:tic245}
\end{figure*}

\begin{figure*}[h]
\begin{center}

\includegraphics[width=0.24\linewidth]{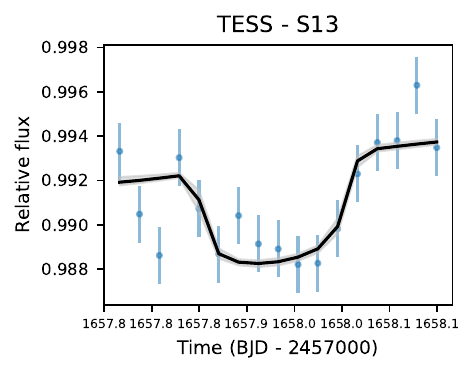}
\includegraphics[width=0.24\linewidth]{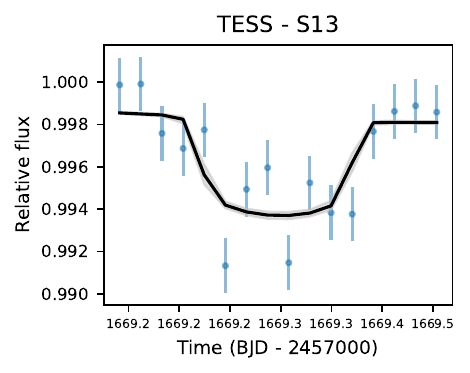}
\includegraphics[width=0.24\linewidth]{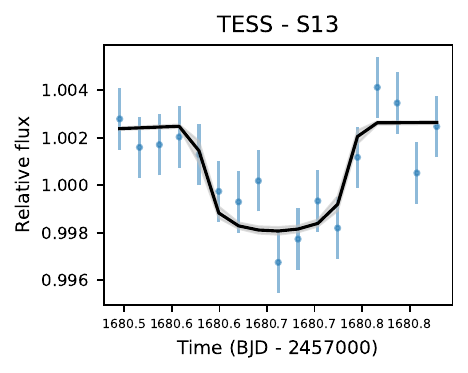}\\
\includegraphics[width=0.24\linewidth]{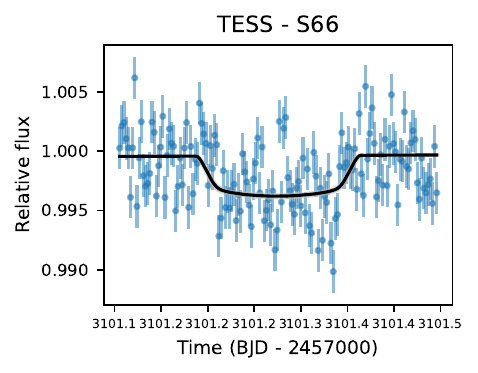}
\includegraphics[width=0.24\linewidth]{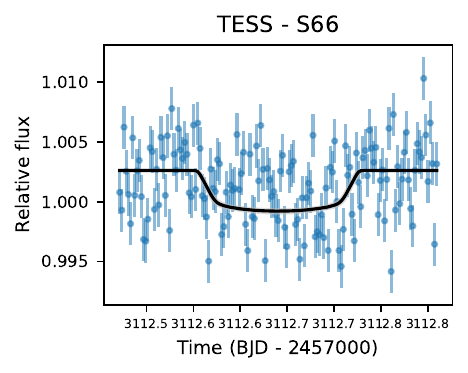}
\includegraphics[width=0.24\linewidth]{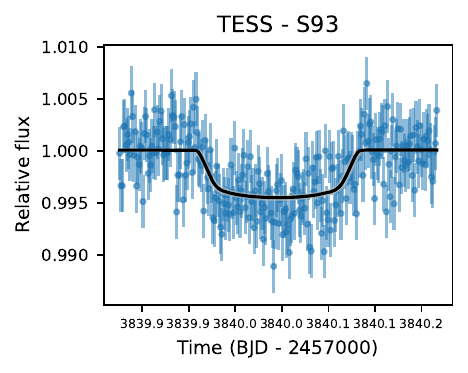}
\includegraphics[width=0.24\linewidth]{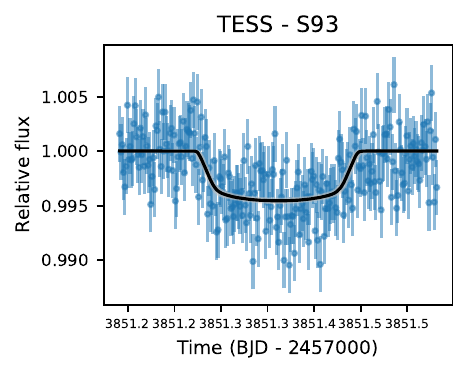}
\includegraphics[width=0.55\linewidth]{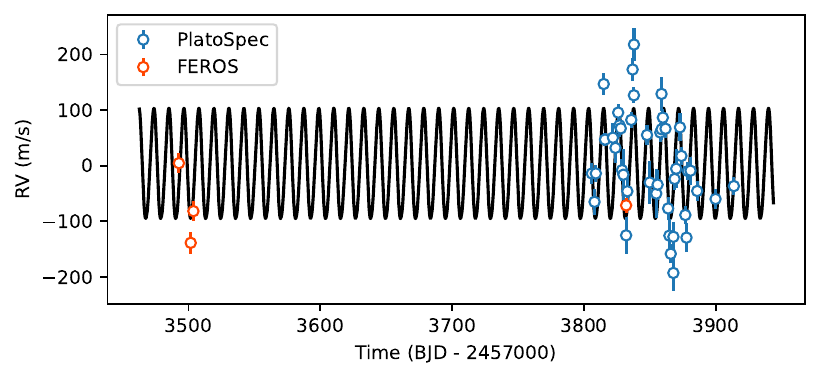}
\includegraphics[width=0.40\linewidth]{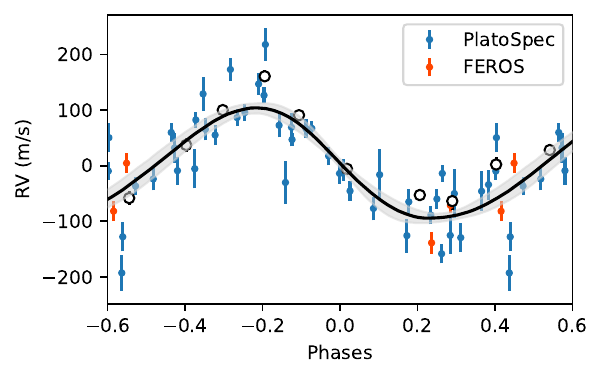}
\end{center}
\caption{Light curves and RV time series of TIC\,87422071. For detail description see Fig.~\ref{fig:tic147}. Binned RV curve is calculated only using PLATOSpec data.}
\label{fig:tic874}
\end{figure*}

The determined planetary parameters of all three systems are listed in Tab.~\ref{tab:planet}. We derived additional parameters from the fitted ones. The derived parameters are orbital inclination $i$, semi-major axis $a$, planetary mass $M_p$ and radius $R_p$, planet density $\rho_p$, surface gravity $g_p$, equilibrium temperature $T_{eq}$, transit depth $\delta$ and durations $T_{14}$ and $T_{23}$. The equilibrium temperature was estimated under the assumption of zero albedo and perfect heat redistribution \citep{Rodriguez2025}. The best-fit model for the light curves of individual TESS sectors and phase-folded RV curves for individual systems are shown in Fig.~\ref{fig:tic147}, \ref{fig:tic245}, and \ref{fig:tic874}. Figures~\ref{fig:tic147_ground} and \ref{fig:tic874_ground} show the model of the ground-based light curves.

According to our analysis, TIC\,14027702b is a warm Jupiter with a mass of $1.09^{+0.07}_{-0.13}$~M$_j$, a radius of $0.97 \pm 0.06$~R$_j$, and an orbital period of 44.4 days. The mean density of the planet is 1.44 g/cm$^3$, and it has a low orbital eccentricity ($e=0.127^{+0.055}_{-0.048}$). The equilibrium temperature is $863^{+26}_{-27}$~K.

TIC\,245076932 is orbited by a warm Jupiter-type planet on a significantly eccentric orbit ($e = 0.428^{+0.022}_{-0.019}$) with a period of 21.6 days. The planet has a slightly lower density of $0.75\pm0.05$ g/cm$^3$. Its radius is $0.97\pm0.05$~R$_J$ and has a mass of $0.51^{+0.04}_{-0.05}$~M$_J$. The equilibrium temperature is similar to the previous one, $845\pm25$~K.

The third discovered warm Jupiter orbits star TIC\,87422071 with an orbital period of 11.4 days on a low-eccentric ($e=0.124^{+0.073}_{-0.060}$) orbit. The radius of the planet is $0.97\pm0.08$~R$_J$ and has a mass of $1.29^{+0.10}_{-0.11}$~M$_J$. The mean density is $1.88^{+0.33}_{-0.28}$ g/cm$^3$, and the equilibrium temperature is $1110\pm33$~K.

\subsection{Interior structure modelling}
\label{interior_analysis}

We perform an interior structure analysis for the three exoplanets. We use the GASTLI interior structure model \citep{acuna21,gastli_software}. GASTLI solves the one-dimensional equations of interior structure for exoplanets under the assumption that the interior is fully convective. The interior is divided into two layers: a heavy-element core composed of a 1:1 rock-water mixture and an envelope. The envelope consists of water (as a proxy for metals) and H/He at a cosmogonic mass ratio. For each of these materials -- H/He, water, and rock -- we employ up-to-date equations of state to calculate the density and entropy. Solving the equations of the interior structure of the planet requires defining the pressure and temperature in the outermost computational layer. To this end, we couple GASTLI to a grid of atmospheric models at a constant pressure of $P_{\rm surf} = 1000$ bar, which is high enough for atmospheric models (including the petitCODE grid we use) to solve for radiative transfer, as opacity tables range from low pressures ($10^{-6}$ bar) to 1000 bar. This grid is the default GASTLI grid computed with the self-consistent 1D atmospheric model petitCODE \citep{molliere15,molliere17} for clear atmospheres in chemical equilibrium. petitCODE uses convective adjustment to determine whether an atmospheric layer is convective or radiative, and calculates self-consistently the Bond albedo to estimate the energy balance at the top of the atmosphere. The GASTLI's input parameters are the planetary mass, the core mass fraction (CMF), the atmospheric metallicity log(M/H) (expressed in solar units), the equilibrium temperature at the null Bond albedo $T_{\rm eq}(A_{\rm B}=0)$, and the internal (or intrinsic) temperature $T_{\rm int}$. The internal temperature is defined as the black-body temperature corresponding to the planet's emission. As output, GASTLI computes the total planetary radius at the transit pressure in optical photometry ($R_{\rm planet}$; 20 mbar) and the age. The latter is calculated by solving the equation of thermal cooling for a luminosity $L = 4 \pi R^{2}_{\rm planet}T^{4}_{\rm int}$. In addition, the atmospheric metallicity log(M/H) is converted to envelope metal mass fraction $Z_{\rm env}$ by the atmospheric grid via chemical equilibrium. The total bulk metal mass fraction, $Z_{\rm planet}$, is estimated to be $Z_{\rm planet} = \rm CMF + (1-\rm CMF) \times Z_{\rm env}$. For more details on the GASTLI interior structure model, see \cite{acuna24}.

We compute a grid for each of the three exoplanets at their respective equilibrium temperatures at the null Bond albedo. The grids span their $\pm5\sigma$ mass interval with steps of 0.05$M_{\rm Jup}$. The CMF ranges from 0 to 0.80 in steps of 0.10; log(M/H) ranges from -2 to 2 in steps of 1.0; and $T_{\rm int}$ ranges from 50 to 350 K in steps of 100 K. We linearly interpolate the grids to obtain the radius and the age for a set of input parameters, which together constitute the forward model in our retrievals. We use the Python package \texttt{emcee} \citep{emceeref} to sample the posterior distribution functions. In our retrievals, the mass, CMF, log(M/H), and $T_{\rm int}$ are free parameters. We adopt a Gaussian prior on the mass, with the mean and standard deviation set to the observed mean and the uncertainty values reported in Table \ref{tab:planet}. The parameters CMF, log(M/H), $T_{\rm int}$ have uniform priors: $\mathcal{U}(0,0.8)$, $\mathcal{U}(-2,2)$, $\mathcal{U}(50,350)$, respectively. The likelihood function is evaluated from the squared residuals of the mass, radius, and age \citep{dorn15,acuna21}. 

\begin{figure}
    \centering
    \includegraphics[width=\columnwidth]{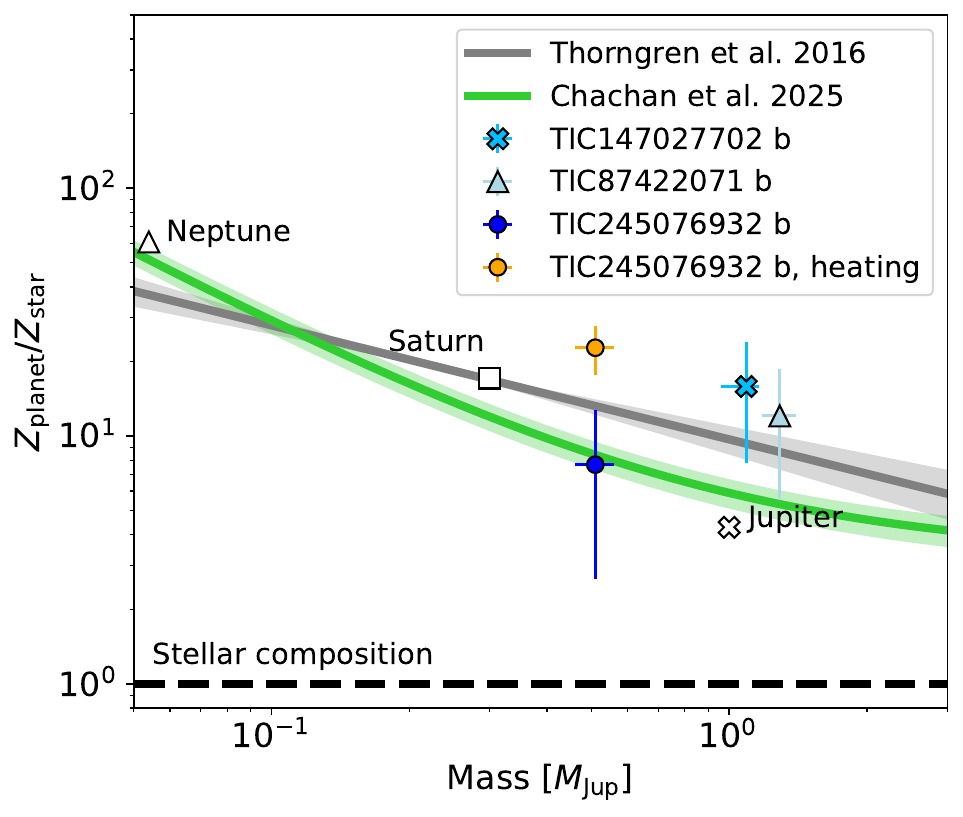}
    \caption{Mass-bulk metal mass fraction diagram for the three exoplanets. We show Neptune, Saturn and Jupiter \citep[][and references therein]{SS_review_interiors} as well as the exoplanet trends fitted by \cite{Thorngren16} and \cite{Chachan}.}
    \label{fig:chachan_diagram}
\end{figure}

Table \ref{tab:int} shows the values retrieved from the compositional parameters for the three exoplanets. We estimate the planet-to-star bulk metal mass fraction ratio by dividing the retrieved $Z_{\rm planet}$ by the stellar metallicity, $Z_{\rm star} = 0.0152\cdot10^{\rm [Fe/H]_{\star}}$ \citep{tala:2025}. Fig. \ref{fig:chachan_diagram} shows the mass-$Z_{\rm planet}/Z_{\rm star}$ diagram of the three exoplanets, including the two exoplanet population trends estimated by \cite{Thorngren16} and \cite{Chachan}. TIC\,245076932\,b and TIC\,87422071\,b are consistent with both trends in 1$\sigma$, while TIC\,147027702\,b is consistent with the fit of \cite{Chachan} in 2$\sigma$.

Additionally, TIC\,245076932\,b has a high eccentricity $e \sim 0.43$. The internal temperature of high-eccentricity exoplanets can be raised above the value predicted by secular cooling due to tidal heating \citep{leconte_tides,millholland_tides}. Radius inflation becomes significant at $T_{\rm eq} > 1000$ K, well above the irradiation level of TIC\,245076932\,b. However, high internal temperatures ($T_{\rm int}\sim$400 K) have been observed in WASP-107\,b and V1298\,Tau\,b through disequilibrium chemistry and transmission spectroscopy \citep{sing_107,welbanks_107,barat25}. These two extrasolar gas giants have lower eccentricities than TIC\,245076932\,b, and it is currently debated whether their additional heating arises from Ohmic dissipation or from super-adiabatic regions induced by magnetic fields and compositional gradients, respectively \citep{barat25,batygin25}. Thus, we consider an alternative scenario in which TIC\,245076932\,b has a high internal temperature $T_{\rm int} = 350$\,K. This scenario illustrates the effect of the heating mechanisms in inference of the compositional parameters. This scenario leads to a high core mass ($M_{\rm core}$) and $Z_{\rm planet}$, as the increase in $T_{\rm int}$ expands the radius of the planet, requiring a larger metal content to fit the observed mass and radius. In this case, we decouple $T_{\rm int}$ from the age by removing the age-squared residuals from the likelihood function in the retrieval.  

Core accretion models predict core masses of at least $\sim 10 \ M_{\oplus}$ \citep{pollack96}. All three exoplanets have core masses above this threshold, consistent with an initial core formation stage followed by runaway gas accretion. However, TIC\,245076932\,b could have a core mass as low as 4\,M${\oplus}$ within 1$\sigma$. A refinement of the stellar age is required to rule out such a low core mass. If this were the case, TIC\,245076932\,b would not be consistent with the core accretion paradigm. Additionally, to determine whether these three exoplanets accreted solids during the runaway accretion phase, atmospheric characterisation would be required to estimate their atmospheric metallicities and carbon-to-oxygen ratios, which are valuable diagnostics of pebble and planetesimal accretion \citep{danti23,feinstein24}. 

\begin{table*}[h]
\centering
\caption{Interior composition parameters retrieved by our interior structure retrievals.}
\label{tab:int}
\resizebox{\textwidth}{!}{%
\begin{tabular}{lcccc}
	\hline
	                                                                             &    TIC\,147027702\,b     &    TIC\,245076932\,b     & TIC\,245076932\,b, heating &      TIC\,87422071       \\ \hline
	Core Mass Fraction, CMF                                                      & 0.16$^{+0.11}_{-0.16}$ & 0.09$^{+0.09}_{-0.06}$ &  0.33$^{+0.08}_{-0.10}$  & 0.21$^{+0.16}_{-0.12}$ \\
	Core Mass, $M_{\rm core}$ [M$_{\oplus}$]                                     &     55.9$\pm$34.0      &     16.4$\pm$12.4      &      50.1$\pm$16.3       &     92.0$\pm$58.5      \\
	Envelope metal mass fraction, $Z_{\rm env}$                                  & 0.01$^{+0.07}_{-0.01}$ & 0.01$^{+0.04}_{-0.01}$ &  0.01$^{+0.10}_{-0.01}$  & 0.01$^{+0.08}_{-0.01}$ \\
	Internal temperature, $T_{\rm int}$ [K]                                      &   141$^{+25}_{-26}$    &    86$^{+11}_{-10}$    &      350 (constant)      &       148$\pm$31       \\
	Bulk metal mass fraction, $Z_{\rm planet}$                                   & 0.19$^{+0.11}_{-0.09}$ & 0.11$^{+0.09}_{-0.07}$ &      0.36$\pm$0.07       & 0.25$^{+0.16}_{-0.12}$ \\
	Planet-to-star bulk metal mass fraction ratio, $Z_{\rm planet}/Z_{\rm star}$ &     15.85$\pm$8.09     &     7.67$\pm$5.02      &      22.73$\pm$5.06      &     12.06$\pm$6.51     \\ \hline
\end{tabular}%
}
\end{table*}

For comparison, we also computed the interior composition of the three planets using modules for experiments in astrophysics \citep[MESA;][]{paxton2011,paxton2013}. For this, we closely followed the implementation presented in \cite{jones:2024} and \cite{tala:2025}, but with some minor modifications. In the case of TIC\,147027702\,b, we adopted the planetary envelope metallicity to be the same as that of the parent star, and we only varied the mass of the inert rocky core, whose density is estimated by assuming a 1:1 mixture of rock and ice \citep{HM1989}. On the other hand, for TIC\,87422071\,b, we increased the metallicity of the planet envelope to match its measured radius with our models, while the prescription to model the rocky core was identical to that for TIC\,147027702\,b. The results for these two planets are presented in Table \ref{tab:int-mesa}. 
Finally, in the case of TIC 245076932\,b, and given its small radius and high irradiation level, we could not find a satisfactory model that reproduces its observed properties.

\begin{table*}[h]
\centering
\caption{Interior composition parameters from MESA.}
\label{tab:int-mesa}
\begin{tabular}{lcc}
	\hline
	                                                                              &      TIC\,147027702\,b      & TIC\,87422071 \\ \hline
	Core Mass Fraction, CMF                                                       &  0.16$^{+0.07}_{-0.09}$   &   0.13$^{+0.13}_{-0.13}$    \\
	Core Mass, $M_{\rm core}$ [M$_{\oplus}$]                                      &  56.0$^{+24.0}_{-32.0}$   &   55.0$^{+55.0}_{-55.0}$   \\
	Envelope metal mass fraction, $Z_{\rm env}$                                   & 0.013 &  0.038   \\
	Core density [g/cm$^3$]                                                       &            9.5            &  11.0     \\
	Bulk metal mass fraction, $Z_{\rm planet}$                                    &  0.17$^{+0.07}_{-0.09}$   & 0.17$^{+0.13}_{-0.13}$   \\
	Planet-to-star bulk metal mass fraction ratio , $Z_{\rm planet}/Z_{\rm star}$ &   13.4$^{+5.5}_{-7.1}$    & 7.6 $^{+5.9}_{-5.9}$ \\ \hline
\end{tabular}
\end{table*}

\section{Discussion and Conclusions}
\label{con}

We report the discovery and characterisation of three warm Jupiter planets orbiting TIC\,14027702, TIC\,245076932 and TIC\,87422071. These are the first exoplanets detected with the new temperature-stabilised spectrograph PLATOSpec \citep{Kabath2025}.
TIC\,147027702b is a Jovian planet with a mass of 1.09~M$_J$ and an orbital period of 45 days. With a time-averaged equilibrium temperature below 900~K, it fits the population of warm Jupiter planets. Planet TIC\,245076932b has a significant eccentric orbit ($e = 0.43$) with an orbital period of 21.6 days. Having a mass of 0.51~M$_J$ and a radius of 0.97~R$_J$, its bulk density is therefore less than 800 kg/m$^3$. The third detected planet, TIC\,87422071b, has a mass of 1.29~M$_J$ and orbits its parent star in 11.4 days. Due to a smaller distance from a star, its equilibrium temperature is slightly above 1100~K, which is still consistent with the population of warm Jupiter planets. 

\begin{figure}
    \centering
    \includegraphics[width=\columnwidth]{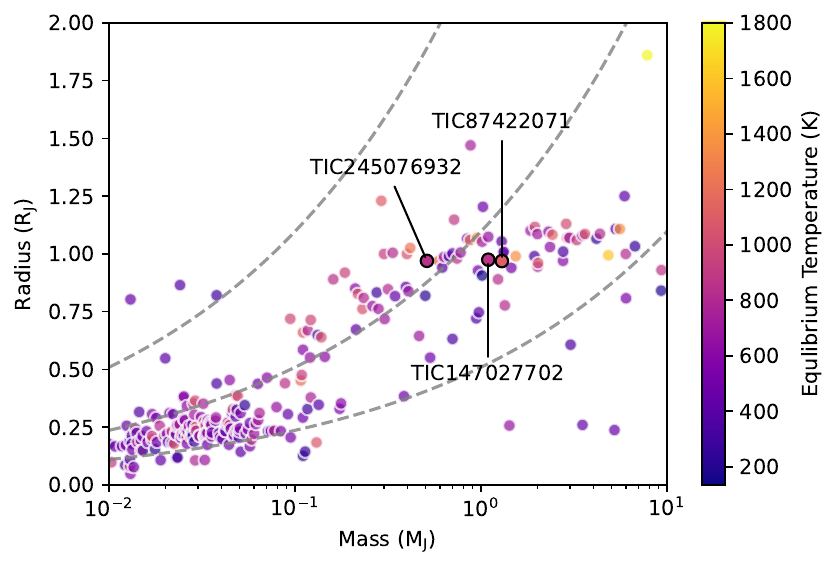}
    \caption{Mass-radius diagram for the population of warm giant planets ($P>10$ days). Dashed lines correspond to bulk densities of 0.1~g/cm$^3$, 1~g/cm$^3$, and 10~g/cm$^3$ (from top to bottom).}
    \label{fig:mass-radius}
\end{figure}

\begin{figure}
    \centering
    \includegraphics[width=\columnwidth]{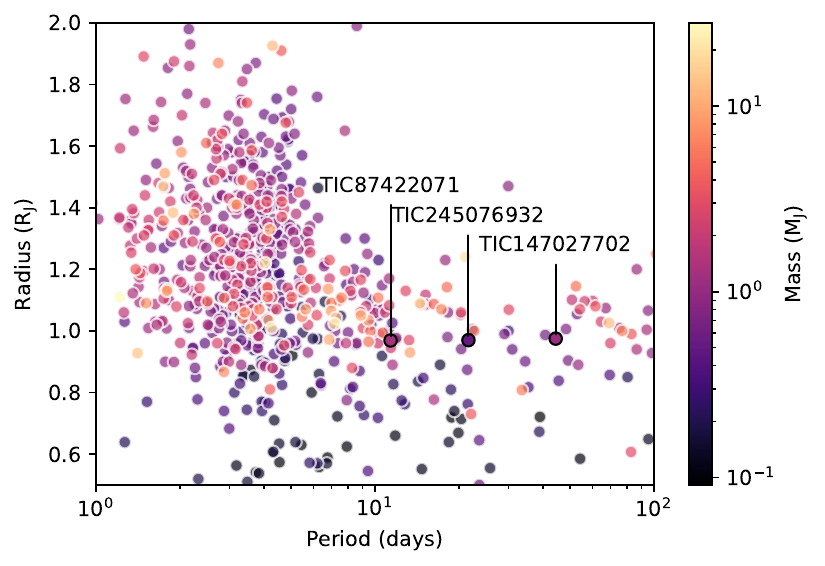}
    \caption{Period-radius diagram for the population of giant planets ($M>0.1$ M$_{\rm J}$).}
    \label{fig:period-radius}
\end{figure}

\begin{figure}
    \centering
    \includegraphics[width=\columnwidth]{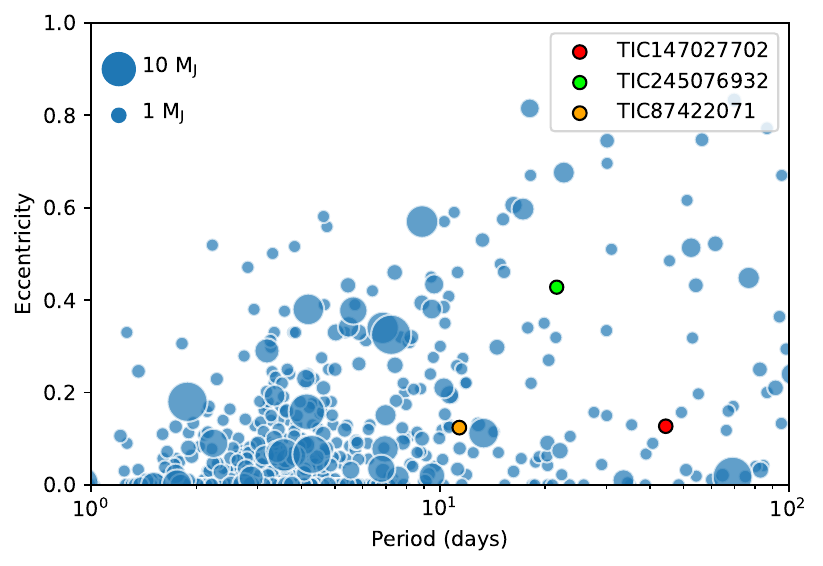}
    \caption{Period-eccentricity diagram for the population of transiting giant planets ($M>0.1$ M$_{\rm J}$).}
    \label{fig:period-ecc}
\end{figure}

Figure \ref{fig:mass-radius} shows the population of warm giant exoplanets in a mass-radius diagram. All three planets presented in this study are uninflated and close to the bulk density level of 1~g/cm$^3$. These new exoplanets lie in a sparsely populated region of the period-radius diagram (Fig.~\ref{fig:period-radius}), corresponding to warm Jupiter planets with orbital periods greater than 10 days. Additionally, TIC\,245076932b has a moderately eccentric orbit (see Fig.~\ref{fig:period-ecc}). The other two planets of this study have nearly circular orbits more frequent in the known sample of warm Jupiters.

The population of warm giant planets is also interesting for performing future atmospheric characterisation \citep[e.g.][]{toi-199}, and linking atmospheric composition with possible formation channels. According to our interior structure modelling, the three planets presented in this study are significantly enriched in heavy elements, and these could translate into high abundances of C, N, and O dominated molecules.
We followed \cite{Kempton2018} to calculate the transmission spectroscopy metric (TSM) and the emission spectroscopy metric (ESM). The TSM values of TIC\,147027702b, TIC\,245076932b, and TIC\,87422071b are $7.3 \pm 3.3$, $35.6 \pm 13.7$ and $12.8 \pm 6.5$, respectively. TIC\,245076932b appears to be the best candidate for transmission spectroscopy, but is still hardly reachable using JWST/NIRISS \citep{niriss}. However, this planet seems to be a suitable target for future instruments such as ARIEL \citep{Ariel} or ELT/ANDES \citep{Andes}. The ESM values are $5.3 \pm 0.9$ (TIC\,147027702b), $13.2 \pm 1.6$ (TIC\,245076932b), and $16.6 \pm 3.0$ (TIC\,87422071b).

A crucial observable for constraining the origin and orbital evolution of warm Jupiters is the inclination of the orbit with respect to the rotation axis of the parent star (stellar obliquity). The projected stellar obliquity can be measured by observing the Rossiter-McLauglin (RM) effect during transit \citep[e.g.][]{Zak2025}. All three stars in the present study are slow rotators. Therefore, the predicted maximal amplitude of the RM effect (i.e. for aligned orbits) is in the range of 18--25 m/s, which would be suitable for instruments like HARPS and ESPRESSO. The significant eccentricity of TIC\,245076932b, makes it an interesting object to measure its stellar obliquity, given that some high-eccentricity migration mechanisms predict significant misalignments \citep{petrovich}. Recent studies targeting moderately eccentric ($\sim 0.3 - 0.6$) warm Jupiters \citep[e.g.][]{espinoza-retamal:2025,Zak2025a}, are showing a tentative trend in the opposite direction, where several mechanisms were proposed to explain the origin of this sub-population of warm Jupiters.

Expanding the sample of well-characterised warm Jupiters, including the three systems presented here, will be essential to disentangle the possible theories of warm- and hot Jupiter planet formation and to place robust constraints on the dominant evolutionary mechanisms at intermediate orbital periods.

\begin{acknowledgements}
We would like to thank L. \v{R}ezba, M. Skarka, and J. \v{S}ubjak for performing observations of these targets. 

This work makes use of observations from the LCOGT network. Part of the LCOGT telescope time was granted by NOIRLab through the Mid-Scale Innovations Program (MSIP). The MSIP is funded by NSF. 
This research has used the website of the Exoplanet Follow-up Observation Programme (ExoFOP; DOI: 10.26134/ExoFOP5), which is operated by the California Institute of Technology, under contract with the National Aeronautics and Space Administration under the Exoplanet Exploration Programme. 
Funding for the TESS mission is provided by NASA's Science Mission Directorate. KAC acknowledges support from the TESS mission via subaward s3449 from MIT.

We would like to acknowledge the dedicated staff of the French and Italian polar agencies (IPEV and PNRA) for their dedication and for their work, particularly those who winter-over, which is essential in maintaining the Concordia Station operational throughout the Austral winter, and thanks to whom the ASTEP$+$ telescope could collect data for this publication. ASTEP$+$ benefited from the support of the French and Italian polar agencies IPEV and PNRA in the framework of the Concordia station programme, from OCA, INSU, Idex UCAJEDI (ANR-15-IDEX-01) and ESA through the Science Faculty of the European Space Research and Technology Centre (ESTEC). ASTEP$+$ also received funding through the Science and Technology Facilities Council (STFC; grants n$^\circ$ ST/S00193X/1, ST/W002582/1 and ST/Y001710/1) as well as the European Research Council (ERC) under the Horizon 2020 research and innovation programme of the European Union (grant agreement n$^\circ$ 803193/BEBOP). 

This work was supported by the Slovak Research and Development Agency under contract No. APVV-24-0160, and by the Ministry of Education, Youth and Sports of the Czech Republic by grant LTT-20015. The research of P.G. was supported by the internal grant No. VVGS-2023-2784 of the P. J. {\v S}af{\'a}rik University in Ko{\v s}ice and funded by the EU NextGenerationEU through the Recovery and Resilience Plan for Slovakia under project No. 09I03-03-V05-00008.

Funding for KB was provided by the European Union (ERC AdG SUBSTELLAR, GA 101054354).

JJ is grateful that publication could be produced within the institutional support framework for the development of the research organisation of Masaryk University.

\end{acknowledgements}

\bibliographystyle{aa}
\bibliography{bibfile}

\begin{appendix}
\section{Radial velocity observations}
\label{ap:rv}

\begin{table}[h]
\caption{Radial-velocity data of TIC\,147027702.}
\label{tab:tic147_rv}
\begin{tabular}{lccc}
	\hline\hline
	Time            &   RV    & RV error & SNR \\
	(BJD)           &  (m/s)  &  (m/s)   &     \\ \hline
	2460686.7921940 & 22707.0 &   36.6   & 28  \\
	2460695.8248905 & 22678.4 &   24.0   & 45  \\
	2460695.8400650 & 22724.9 &   21.1   & 52  \\
	2460701.8156190 & 22764.0 &   22.1   & 49  \\
	2460702.8331103 & 22740.6 &   27.5   & 39  \\
	2460704.8489000 & 22816.8 &   27.4   & 39  \\
	2460710.8349756 & 22866.1 &   31.7   & 34  \\
	2460717.8033857 & 22724.9 &   38.7   & 27  \\
	2460721.8378543 & 22765.9 &   31.0   & 34  \\
	2460724.8093707 & 22757.4 &   20.3   & 54  \\
	2460731.7910830 & 22726.8 &   35.7   & 29  \\
	2460736.8218505 & 22680.3 &   21.9   & 50  \\
	2460738.8154537 & 22711.0 &   20.2   & 55  \\
	2460742.8102396 & 22751.7 &   30.0   & 35  \\
	2460743.7871938 & 22674.6 &   21.2   & 52  \\
	2460744.7619825 & 22753.9 &   22.0   & 50  \\
	2460745.7432980 & 22726.1 &   19.7   & 56  \\
	2460749.7804488 & 22785.0 &   25.8   & 43  \\
	2460751.8135311 & 22816.5 &   23.9   & 46  \\
	2460753.7393160 & 22780.7 &   21.2   & 52  \\
	2460757.7221696 & 22791.0 &   20.9   & 52  \\
	2460758.6920467 & 22773.2 &   41.7   & 25  \\
	2460760.6354317 & 22755.8 &   32.3   & 33  \\
	2460766.7161256 & 22826.8 &   18.8   & 59  \\
	2460768.7003831 & 22748.1 &   20.5   & 54  \\
	2460774.6598235 & 22682.9 &   30.2   & 36  \\
	2460778.7205240 & 22662.2 &   26.0   & 42  \\
	2460781.7031698 & 22708.6 &   19.3   & 58  \\
	2460783.6313187 & 22690.0 &   22.6   & 47  \\
	2460786.6357072 & 22674.1 &   35.3   & 30  \\
	2460786.6612736 & 22715.9 &   29.5   & 37  \\
	2460788.6784084 & 22748.2 &   19.3   & 58  \\
	2460790.6384807 & 22722.5 &   20.1   & 55  \\
	2460791.6160357 & 22786.9 &   21.1   & 51  \\
	2460792.6242137 & 22783.7 &   32.9   & 32  \\
	2460794.6346524 & 22688.5 &   23.1   & 47  \\
	2460797.6539720 & 22752.6 &   25.9   & 42  \\
	2460799.6221343 & 22780.0 &   27.2   & 34  \\
	2460798.6870562 & 22820.1 &   31.6   & 39  \\
	2460800.5909948 & 22732.4 &   25.7   & 41  \\
	2460805.6333315 & 22870.6 &   33.3   & 32  \\
	2460806.6896655 & 22744.1 &   27.6   & 40  \\
	2460807.5668869 & 22817.9 &   25.1   & 43  \\
	2460808.6134686 & 22774.7 &   23.4   & 47  \\
	2460809.5378564 & 22819.7 &   16.4   & 70  \\
	2460810.6896164 & 22697.5 &   31.7   & 35  \\
	2460815.6550418 & 22706.6 &   26.3   & 41  \\
	2460819.5596421 & 22741.3 &   27.8   & 38  \\
	2460823.5642501 & 22793.4 &   35.6   & 30  \\
	2460824.5241531 & 22704.4 &   19.5   & 57  \\
	2460825.6729563 & 22770.9 &   27.8   & 42  \\
	2460837.5533146 & 22785.3 &   25.6   & 43  \\
	2460848.5297432 & 22770.6 &   30.6   & 35  \\
	2460858.5583752 & 22669.7 &   30.8   & 36  \\
	2460865.5212450 & 22705.0 &   33.8   & 32  \\ \hline
\end{tabular}          
\end{table}

\begin{table*}[h]
\caption{Radial-velocity data of TIC\,245076932.}
\label{tab:tic245_rv}
\centering  
{
\begin{tabular}{lccc}
	\hline\hline
	Time            &   RV    & RV error & SNR \\
	(BJD)           &  (m/s)  &  (m/s)   &     \\ \hline
	2460711.8505993 & -4114.7 &   16.2   & 31  \\
	2460721.8684073 & -4173.0 &   11.9   & 42  \\
	2460724.8308937 & -4166.9 &   11.5   & 43  \\
	2460733.7977187 & -4071.0 &   19.4   & 25  \\
	2460737.8368879 & -4141.0 &   12.0   & 41  \\
	2460739.8316922 & -4154.3 &   11.8   & 42  \\
	2460740.8550901 & -4173.4 &   15.1   & 33  \\
	2460742.8645549 & -4195.2 &   13.9   & 36  \\
	2460743.8506680 & -4183.0 &   13.4   & 37  \\
	2460744.8476837 & -4172.5 &   15.9   & 31  \\
	2460745.8530324 & -4156.3 &   11.1   & 45  \\
	2460746.8913739 & -4184.8 &   11.8   & 42  \\
	2460747.8593865 & -4196.9 &   10.4   & 49  \\
	2460749.8505336 & -4187.7 &   13.0   & 38  \\
	2460750.8608490 & -4229.1 &   13.6   & 37  \\
	2460751.9015180 & -4187.9 &   16.8   & 30  \\
	2460753.8357969 & -4178.8 &   11.8   & 42  \\
	2460754.8381103 & -4133.5 &   13.6   & 36  \\
	2460756.7853991 & -4142.7 &   12.3   & 40  \\
	2460757.7892431 & -4150.4 &   13.8   & 35  \\
	2460758.7420421 & -4163.3 &   23.3   & 21  \\
	2460761.7903716 & -4184.9 &   24.8   & 21  \\
	2460761.8182084 & -4162.6 &   30.3   & 17  \\
	2460766.8173540 & -4168.0 &   16.4   & 30  \\
	2460767.7971582 & -4165.6 &   21.1   & 24  \\
	2460774.7592066 & -4199.1 &   17.1   & 28  \\
	2460777.7719106 & -4066.5 &   25.4   & 21  \\
	2460777.7938556 & -4089.3 &   24.6   & 21  \\
	2460778.8181846 & -4129.4 &   19.6   & 27  \\
	2460778.8469700 & -4096.3 &   20.7   & 26  \\
	2460779.8045949 & -4105.4 &   17.0   & 31  \\
	2460779.8324427 & -4071.9 &   19.6   & 27  \\
	2460781.7800507 & -4166.3 &   11.3   & 45  \\
	2460783.7382212 & -4148.8 &   16.6   & 30  \\
	2460786.7501504 & -4175.0 &   14.1   & 36  \\
	2460788.8115154 & -4133.0 &   14.2   & 35  \\
	2460790.7512885 & -4187.8 &   13.5   & 37  \\
	2460791.7542462 & -4213.0 &   13.5   & 37  \\
	2460792.7106400 & -4153.1 &   22.2   & 23  \\
	2460793.8391743 & -4189.5 &   12.3   & 42  \\ \hline
\end{tabular}
\hspace{1cm}
\begin{tabular}{lccc}
	\hline\hline
	Time            &   RV    & RV error & SNR \\
	(BJD)           &  (m/s)  &  (m/s)   &     \\ \hline
	2460794.7420006 & -4213.4 &   12.3   & 40  \\
	2460797.7424400 & -4189.2 &   10.3   & 50  \\
	2460798.7707242 & -4135.7 &   15.7   & 33  \\
	2460799.7252226 & -4140.3 &   09.4   & 55  \\
	2460800.6578914 & -4110.3 &   12.2   & 41  \\
	2460802.7360130 & -4161.6 &   14.8   & 34  \\
	2460805.7193731 & -4125.4 &   12.5   & 40  \\
	2460806.7745561 & -4181.4 &   17.6   & 30  \\
	2460807.6433633 & -4177.4 &   11.6   & 43  \\
	2460808.6857083 & -4198.4 &   10.2   & 51  \\
	2460814.7840132 & -4245.7 &   17.0   & 31  \\
	2460815.7354949 & -4211.1 &   10.6   & 49  \\
	2460816.7158521 & -4191.4 &   10.8   & 47  \\
	2460819.6696939 & -4169.7 &   18.2   & 28  \\
	2460820.4827714 & -4148.5 &   14.8   & 34  \\
	2460821.6545058 & -4144.9 &   15.3   & 33  \\
	2460821.6828498 & -4104.3 &   13.4   & 38  \\
	2460823.6058421 & -4170.7 &   16.6   & 30  \\
	2460823.6349615 & -4111.7 &   17.9   & 28  \\
	2460824.6128327 & -4203.6 &   11.6   & 44  \\
	2460825.7411024 & -4171.7 &   11.8   & 44  \\
	2460828.5918274 & -4161.3 &   11.0   & 46  \\
	2460829.6110110 & -4138.6 &   23.5   & 21  \\
	2460835.5910153 & -4164.5 &   10.8   & 48  \\
	2460836.6383962 & -4187.1 &   10.8   & 48  \\
	2460837.6614717 & -4209.5 &   11.6   & 45  \\
	2460847.6477437 & -4153.3 &   14.5   & 35  \\
	2460854.6321937 & -4135.9 &   45.7   & 12  \\
	2460863.5972044 & -4130.9 &   14.1   & 36  \\
	2460868.5748781 & -4173.9 &   22.5   & 23  \\
	2460869.6238660 & -4145.9 &   36.8   & 16  \\
	2460872.5404851 & -4254.2 &   23.3   & 21  \\
	2460874.5620393 & -4183.2 &   14.0   & 36  \\
	2460876.6452774 & -4255.0 &   18.4   & 30  \\
	2460878.6032402 & -4163.2 &   25.1   & 21  \\
	2460880.5676253 & -4167.3 &   19.1   & 26  \\
	2460885.5648661 & -4135.3 &   19.5   & 27  \\
	2460895.5209800 & -4232.4 &   41.5   & 13  \\
	2460901.5866744 & -4195.2 &   21.0   & 27  \\ \hline
	                &         &          &
\end{tabular} 	}
\end{table*}	

\begin{table}[h]
\caption{Radial-velocity data of TIC\,87422071 obtained by PLATOSpec.}
\label{tab:tic874_rv}
\begin{tabular}{lccc}
	\hline\hline
	Time            &   RV   & RV error & SNR \\
	(BJD)           & (m/s)  &  (m/s)   &     \\ \hline
	2460805.8834210 & 2745.7 &   18.7   & 35  \\
	2460807.9155147 & 2694.7 &   22.8   & 29  \\
	2460808.8980161 & 2745.8 &   14.9   & 47  \\
	2460814.8751567 & 2906.6 &   19.6   & 34  \\
	2460815.8557630 & 2806.6 &   12.1   & 58  \\
	2460821.8501538 & 2810.1 &   26.2   & 26  \\
	2460823.7741806 & 2791.7 &   26.7   & 25  \\
	2460825.8367722 & 2855.2 &   15.2   & 44  \\
	2460826.8410798 & 2832.3 &   21.7   & 31  \\
	2460827.8000835 & 2826.8 &   12.7   & 54  \\
	2460828.7406824 & 2751.7 &   19.4   & 35  \\
	2460829.7904270 & 2743.5 &   46.0   & 16  \\
	2460831.8612531 & 2634.4 &   33.1   & 20  \\
	2460832.7799130 & 2713.7 &   35.6   & 19  \\
	2460835.7637607 & 2842.0 &   14.2   & 48  \\
	2460836.7780307 & 2932.7 &   20.9   & 32  \\
	2460837.7644513 & 2886.3 &   14.2   & 47  \\
	2460837.8028896 & 2977.6 &   29.5   & 23  \\
	2460847.6944565 & 2814.9 &   18.1   & 37  \\
	2460849.7573677 & 2729.5 &   37.9   & 18  \\
	2460854.7100868 & 2709.6 &   44.8   & 16  \\
	2460855.7134477 & 2725.6 &   27.2   & 25  \\
	2460857.7724212 & 2819.4 &   17.2   & 39  \\
	2460858.7063917 & 2888.8 &   31.0   & 22  \\
	2460858.7736015 & 2825.1 &   20.1   & 34  \\
	2460859.7045034 & 2846.2 &   15.2   & 45  \\
	2460861.7244720 & 2826.2 &   17.1   & 39  \\
	2460863.7033588 & 2682.8 &   21.4   & 32  \\
	2460864.6837203 & 2633.9 &   30.4   & 24  \\
	2460865.7026675 & 2601.3 &   17.1   & 40  \\
	2460867.6868202 & 2566.8 &   31.8   & 24  \\
	2460867.7146551 & 2631.8 &   26.0   & 27  \\
	2460868.6095502 & 2735.8 &   20.4   & 34  \\
	2460869.8269116 & 2754.2 &   35.1   & 22  \\
	2460872.6654226 & 2828.7 &   25.2   & 27  \\
	2460873.7521680 & 2777.2 &   16.8   & 40  \\
	2460876.7511659 & 2671.0 &   16.7   & 41  \\
	2460877.6283765 & 2630.3 &   25.9   & 26  \\
	2460878.6568412 & 2750.0 &   15.6   & 44  \\
	2460880.6843347 & 2750.7 &   25.5   & 26  \\
	2460885.7461869 & 2714.2 &   17.4   & 39  \\
	2460899.6580413 & 2699.8 &   17.4   & 38  \\
    2460913.5513713 & 2723.0 &   16.7   & 41  \\ \hline
\end{tabular}          
\end{table}

\begin{table}[h]
\caption{Radial-velocity data of TIC\,87422071 obtained by FEROS.}
\label{tab:tic874_rv-feros}
\begin{tabular}{lccc}
	\hline\hline
	Time            &   RV   & RV error & SNR \\
	(BJD)           & (m/s)  &  (m/s)   &     \\ \hline
	2460492.7845381 & 2740.2 &   17.8   &  45   \\
	2460501.7300205 & 2596.9 &   19.4   &  41   \\
	2460503.7693251 & 2653.9 &   17.9   &  45   \\
	2460831.8640888 & 2664.3 &   13.7   &  61   \\ \hline
\end{tabular}          
\end{table}

\section{Activity indices}
\begin{figure}[h]
    \centering
    \includegraphics[width=\columnwidth]{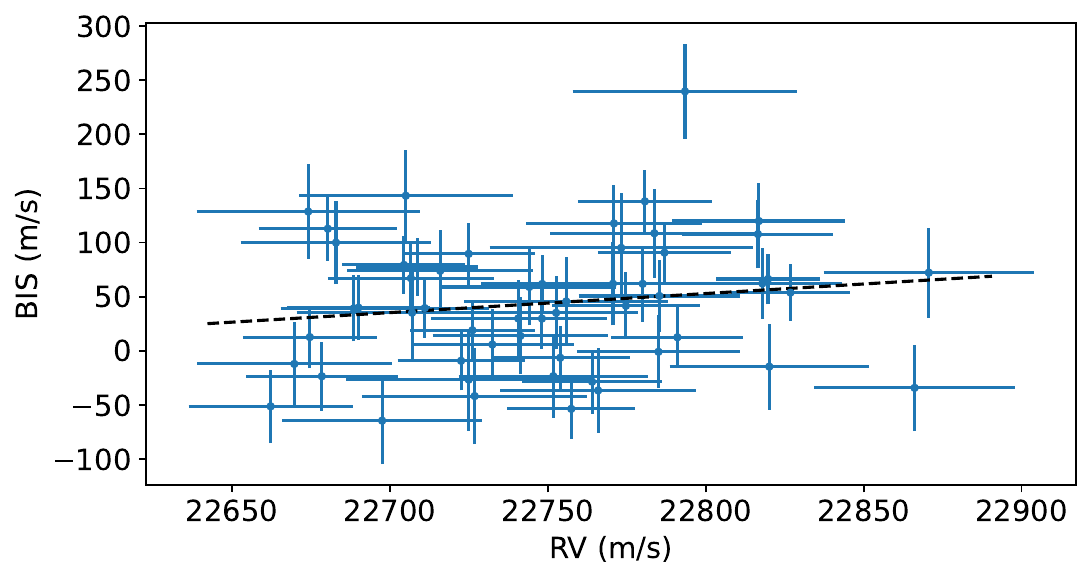}
    \includegraphics[width=\columnwidth]{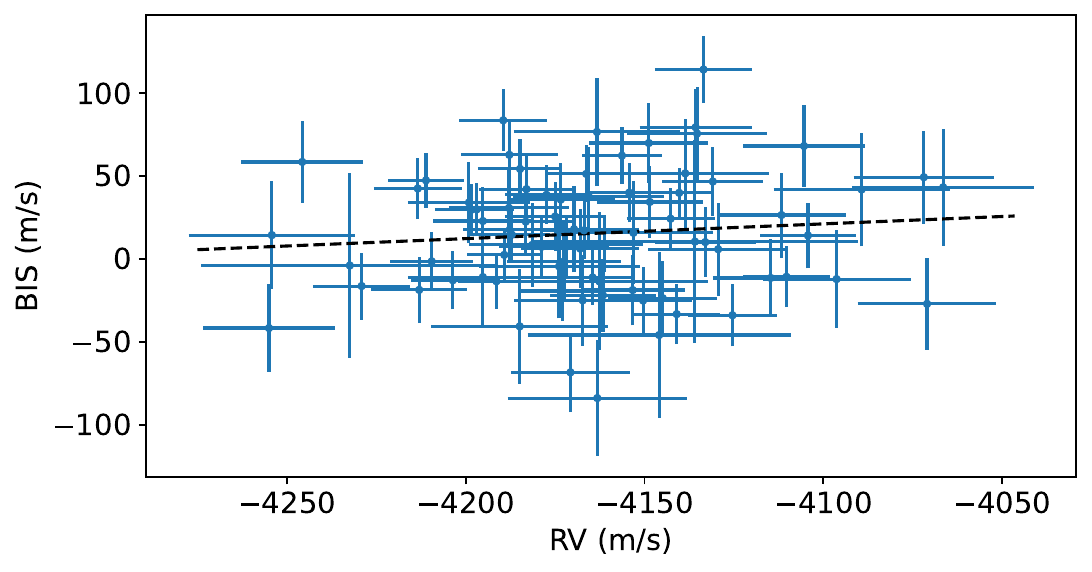}
    \includegraphics[width=\columnwidth]{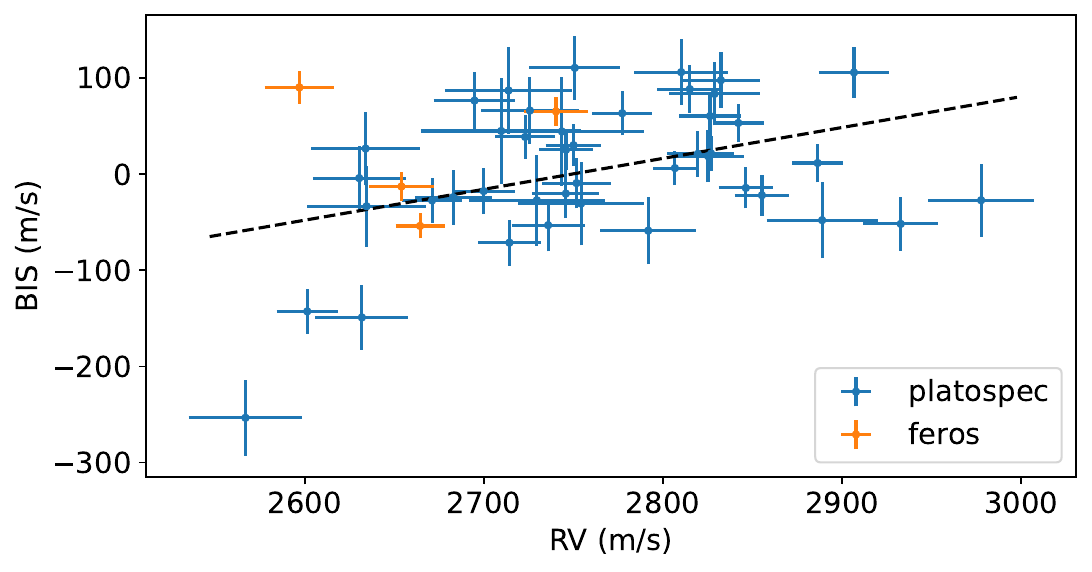}
    \caption{Bisector span (BIS) as a function of RV of TIC\,147027702 (\textit{top}), TIC\,245076932 (\textit{middle}) and TIC\,87422071 (\textit{bottom}). The dashed line shows the best linear fit to the data.}
    \label{fig:bis}
\end{figure}

\section{Corner plots}
\begin{figure*}[h]
\includegraphics[width=\textwidth]{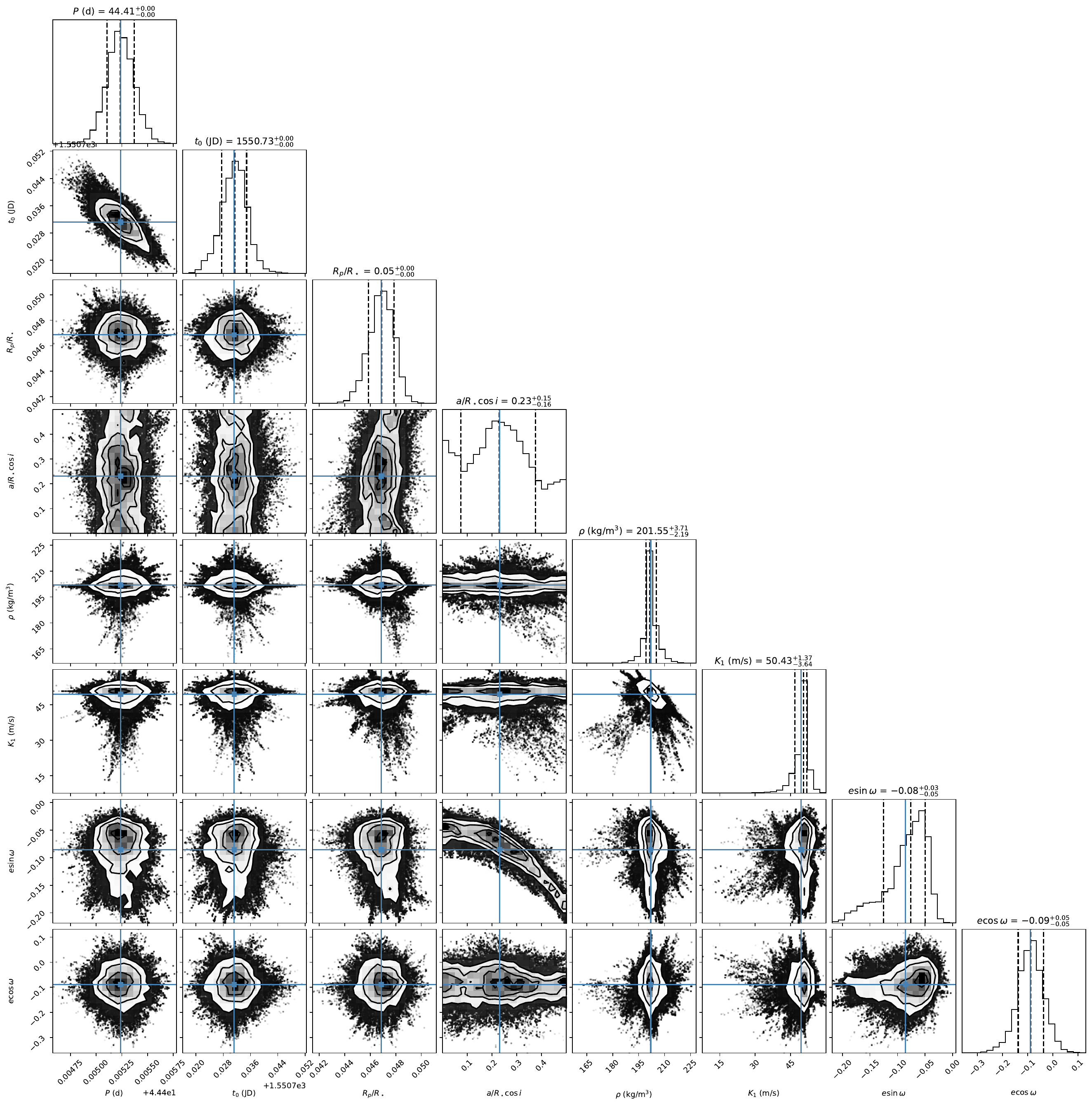}
\caption{Corner plot showing posterior distributions and covariances for some fitted parameters in our analysis of TIC\,147027702.}
\end{figure*}

\begin{figure*}[h]
\includegraphics[width=\textwidth]{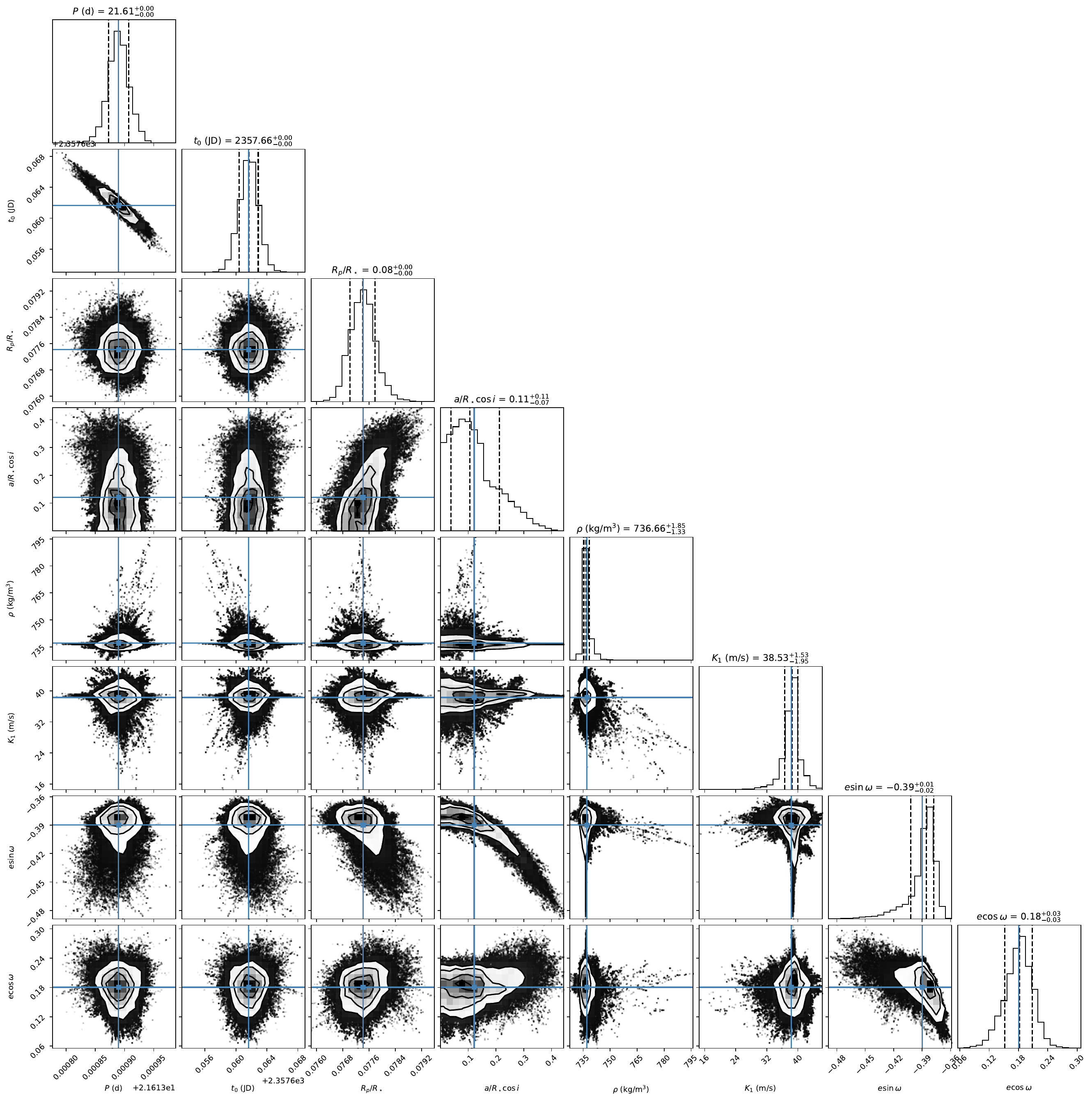}
\caption{Corner plot showing posterior distributions and covariances for some fitted parameters in our analysis of TIC\,245076932.}
\end{figure*}

\begin{figure*}[h]
\includegraphics[width=\textwidth]{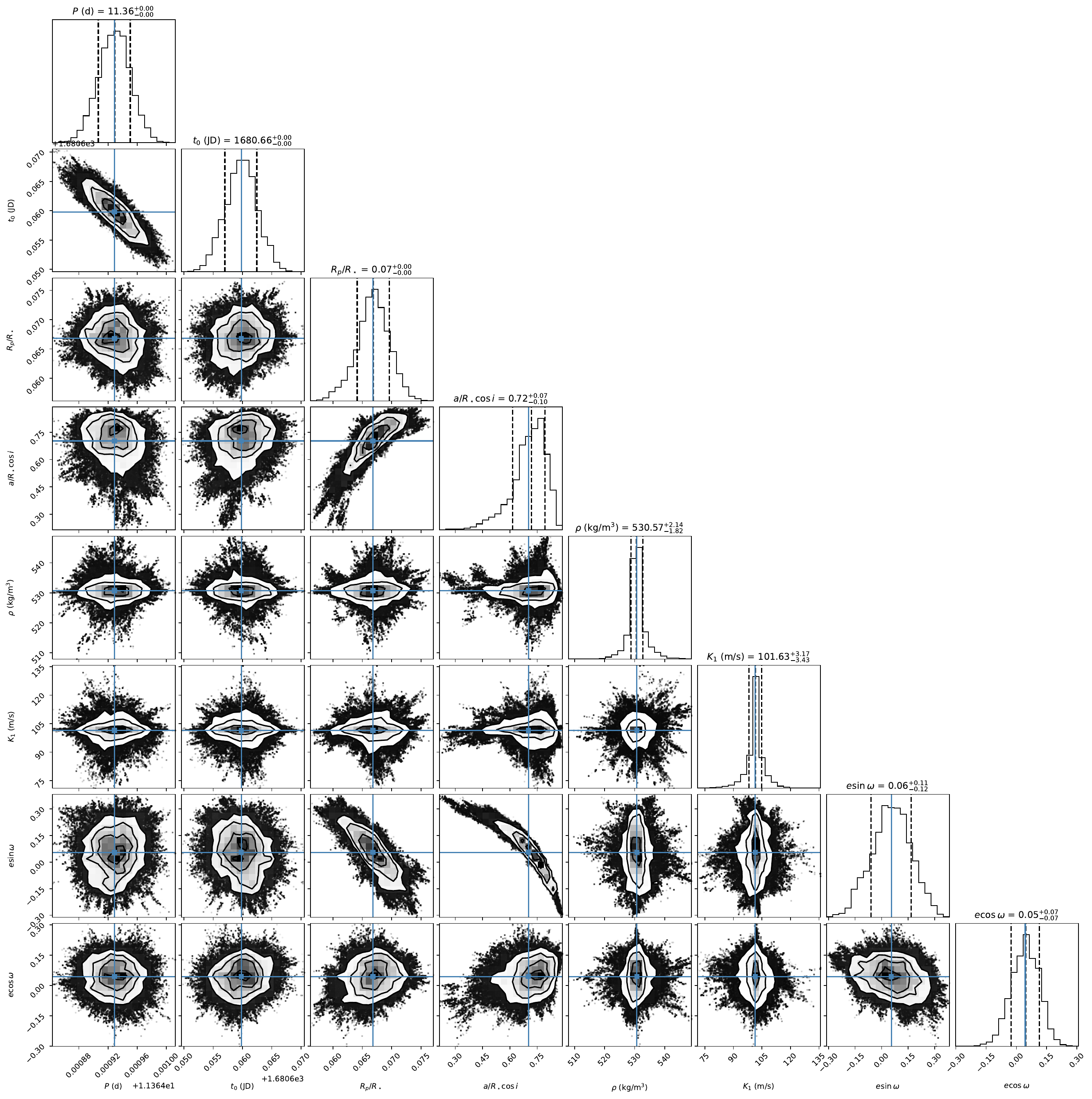}
\caption{Corner plot showing posterior distributions and covariances for some fitted parameters in our analysis of TIC\,87422071.}
\end{figure*}

\end{appendix}

\end{document}